\documentclass[preprint2,10pt,column]{aastex}

\usepackage[
  a4paper, mag=1000, includefoot,
  left=2.0cm, right=1cm, top=2cm, bottom=2cm, headsep=1cm, footskip=1cm
]{geometry}
\usepackage[T2A]{fontenc}
\usepackage[utf8]{inputenc}
\usepackage[english]{babel}
\usepackage{babelbib}
\usepackage{rotating}
\usepackage{floatrow}
\usepackage{longtable}
\usepackage{multirow}
\usepackage{lscape}
\usepackage{array}
\usepackage{cleveref}
\usepackage{graphicx}
\usepackage{natbib}

\ifpdf\usepackage{epstopdf}\fi
\usepackage{journals}

\def\apjl{Astrophys. J. Let}
\def\apjs{Astrophys. J. Suppl}
\def\aap{Astron. and Astrophys}

\def\gee{ \, \lower 1mm\hbox{$\,{\buildrel > \over{\scriptstyle\scriptstyle\sim} }\displaystyle \,$}}
\def\lee{ \, \lower 1mm\hbox{$\,{\buildrel < \over{\scriptstyle\scriptstyle\sim} }\displaystyle \,$}}
\def\ltsima{$\; \buildrel < \over \sim \;$}
\def\simlt{\lower.5ex\hbox{\ltsima}}
\def\gtsima{$\; \buildrel > \over \sim \;$}
\def\simgt{\lower.5ex\hbox{\gtsima}}
\def\etal{{\it et~al.~}}

\def\H2{H$_{2}$}

\def\roH2{$\rho_{\mbox{\footnotesize \H2}}$}

\def\MH2{M$_{\textrm{H}_2}$}

\shorttitle{Stellar populations in isolated S0's}
\shortauthors{I.Yu.~Katkov\etal}

\begin{document}

\title{Properties of Stellar Populations in Isolated Lenticular Galaxies}

\author{Ivan Yu. Katkov}
\affil{Faculty of Physics, Sternberg Astronomical Institute \\Lomonosov Moscow State University
Moscow, 119991, Russia}
\email{katkov.ivan@gmail.com} 

\author{Olga K. Sil'chenko}
\affil{Sternberg Astronomical Institute Lomonosov Moscow State University, Moscow, 119991, Russia}
\email{sil@sai.msu.su} 

\author{Victor L. Afanasiev}
\affil{Special Astrophysical Observatory of the Russian Academy
of Sciences, Nizhnji Arkhyz, 369167 Russia}
\email{vafan@sao.ru} 






\begin{abstract}
In this paper we present the results of long-slit spectral observations for a sample 
of isolated lenticular galaxies, made with the SCORPIO and SCORPIO-2 spectrographs of the 6-meter BTA 
telescope of the SAO RAS. By applying full spectral fitting technique using the stellar population
evolutionary synthesis models, we have measured the radial profiles of the stellar line-of-sight velocity 
as well as the velocity dispersion, SSP-equivalent age and SSP-equivalent metallicity of stars along the radius 
in 12 targets. The resulting averaged ages of the stellar population in bulges and discs  cover an entire range 
of possible values from 1.5 to 15~Gyr which indicates the absence of a certain formation epoch for the
structural components in the isolated lenticular galaxies, unlike in the members of clusters and rich groups: 
they could have been formed at a redshift of $z>2$ as well as only a few billion years ago. Unlike S0 galaxies 
in more dense environments, the isolated galaxies typically have the same age of stars in the bulges and discs. 
The disc-embedded lenses and rings of increased stellar brightness, identified from the photometry in 7 of 11 galaxies, 
do not differ strongly from the stellar discs as concerning the properties of stellar populations and stellar velocity 
dispersion. We conclude that the final shaping of the morphological type of a lenticular galaxy in complete isolation is
critically dependent on the possible regimes of cold-gas accretion from outside.
\end{abstract}

\bigskip


\section{INTRODUCTION}

The problem of formation and evolution paths of galaxies is the
key point in modern extragalactic astronomy and observational cosmology. 
Galaxies are formed being shaped under a large number of physical factors which are 
often poorly known. The main issue here is to identify the most important factors which 
are crucial during the formation and evolution of galaxies of a given
morphological type.

The morphological type of lenticular (S0) galaxies was proposed by Edwin Hubble
as a hypothetical one in 1936~\cite{hubble_1936} in order to
fill in the intermediate position between elliptical and
spiral galaxies. It was assumed that the objects of this type had to have
large-scale stellar discs as observed in the spiral galaxies but
could not demonstrate any noticeable starforming regions and spiral patterns
in their stellar discs. Their smooth reddish view and probably an
old mean age of stars made them to look similar to the elliptical galaxies.
An intermediate position of lenticular galaxies between the purely
spheroidal stellar systems and the spiral galaxies, where the
contribution of the bulge into the total luminosity decreases monotonically 
with the transition from early to late types (from left to right along the
Hubble sequence of morphological types), gives rise to a natural assumption
that the S0 galaxies should possess large bulges. However, the
detailed surface photometry of S0 galaxies has shown
that the bulges in them can be either very large or
tiny \citep{laurikainen_2010}. Stemming from these results, an old idea of
Sidney \citet{van_den_bergh_1976}, proposing that the
lenticular galaxies should be forming a sequence parallel to the
spiral galaxies in the Hubble fork, and the connection between
the (close) position in this diagram of the S0(a, b, c) and the
spiral galaxies of the corresponding subtypes is determined by the
``bulge/disc'' luminosity ratio \citep{kormendy_bender_2012,cappellari_atlas7_2011},
becomes increasingly popular. It would seem that such a turn in
the understanding of the Hubble sequence would only reinforce the conventional 
point of view about the formation of lenticular galaxies via the cessation 
of star formation in the discs of spiral galaxies: the event of the
transformation of a progenitor galaxy into the resulting S0 galaxy is much 
easier to imagine when both galaxies have the same bulge-to-disc ratio.
However, it should be noted that if the contribution of the bulge
to the total luminosity in an S0 galaxy is the same as in the
spiral galaxy located nearby in the morphological diagram, then
this leaves open the possibility of reverse transformation, namely,
the transformation of an S0 galaxy into a spiral one, by gas acquisition and
subsequent star formation burning, which would have been quite impossible in
the presence of a systematically larger bulges in the S0s. 


The literature discusses a huge number of physical
processes that could cease star formation in a disc of a
spiral galaxy. Here there are some of them listed:
\begin{itemize}
\item direct collisions of galaxies \citep{spitzer_1951,icke_1985};
\item gravitational tidal effects from the dark halo of the cluster/group \citep{byrd_1990};
\item ``harassment'' or the gravitational tidal influence
of galaxies on each other during a sufficiently fast flyby \citep{moore_1996};
\item ram pressure of the hot intracluster
medium \citep{gunn_gott_1972,quilis_2000};
\item ``starvation'' or  termination of star formation because of
disappearance of external reservoirs of gas,
previously maintaining the gas accretion onto the disc of a
galaxy and feeding its star formation \citep{larson_1980}.
\end{itemize}

These processes are closely related to the density of environments of
galaxies, because only the clusters and rich groups of galaxies
with their massive dark haloes can provide the necessary density
of the hot intergalactic medium for the ram pressure and tight
packing of galaxies for the efficiency of gravitational tides.

It is known that lenticular galaxies are the dominant population of
nearby clusters of galaxies, where their fraction reaches up
to~60\% \citep{dressler_1980}. However, the number of S0
galaxies is quite noticeable among the field galaxies as well:
according to the APM survey \citep{naim_1995}, the
fraction of lenticular galaxies in the nearby Universe is
about~15\%, and they are the second class by the frequency of occurrence
after the spirals. Furthermore, we know cases of completely
isolated lenticular galaxies \citep{sulentic_2006}. Here we pose
a question which has not yet been raised, about the origin
of such galaxies. Under which physical mechanisms
were they formed, and how do these mechanisms differ from those
that work in the dense environment?

Despite the apparent deficit of mechanisms which may be responsible for the
morphological transformation of isolated galaxies, in comparison to the
members of clusters and groups, it is wrong to suppose that the
isolated galaxies evolve completely internally, following the
``closed box'' scenario. In our work \citep{sil_2011_n7217} we have demonstrated 
that a completely isolated early-type spiral galaxy NGC\,7217 over the
last 5~Gyr experienced at least two events of minor merging. In
a locally isolated S0 galaxy NGC\,4124 we have also found
signatures of minor merging, which had apparently taken place 2--3~Gyr
ago and had provoked a central starburst \citep{zasov_2013_n4124}. 
Furthermore, it was recently understood that lenticular field galaxies 
often possess significant amounts of gas. Moreover, it is exactly in the
rarefied environment that a galaxy often reveals decoupled
kinematics of stars and gas, implying external origin of the
gas \citep{davis_2011}. Thus, the study of the properties of isolated lenticular galaxies 
would allow to concentrate on the evolutionary mechanisms, related  either to
the internal disc instabilities or just the external accretion of
gas and/or companion gas-rich galaxies. It should be noted that the accretion 
of gas and/or minor merging events may not only suppress the star
formation in the disc but, on the contrary, stimulate
its burst \citep{birnboim07,sancisi_rev}.

The targets studied in this work are the nearby, strongly
isolated lenticular galaxies, for which we have fulfiled long-slit
spectroscopic observations with the aim to determine the resolved
properties of stellar populations. The properties of the ionized gas and its
kinematics in comparison to the stellar component kinematics are discussed 
in the separate paper \citep{ilg_gas}, while in this paper we concentrate only 
on the ages and chemical composition of the stellar populations.

The paper is organized as follows: Section~\ref{sample:Katkov_n}
is devoted to the description of the sample; the details of the spectral 
observations, reduction, and data analysis are discussed in
Section~\ref{observations:Katkov_n};
Section~\ref{results:Katkov_n} presents the results for every galaxy; 
Sections~\ref{discussion:Katkov_n}
and~\ref{conclusions:Katkov_n} contain the discussion of the
results and the summary.

\section{SAMPLE SELECTION}\label{sample:Katkov_n}

We have compiled a  sample of nearby isolated lenticular galaxies by using an approach that 
has recently been developped in the Laboratory of Extragalactic Astrophysics and Cosmology of the Special
Astrophysical Observatory (SAO) RAS by Karachentsev, Makarov, and their co-authors. This approach
has been applied by them to the galaxies of the Local Supercluster and its surroundings to extract
pairs \citep{karachentsev_pairs_2008}, triplets \citep{makarov_triplets_2009}, and groups \citep{makarov_groups_2011}, 
as well as to select isolated galaxies \citep{karachentsev_isol_2011}. The data on the line-of-sight (LOS) velocities,   
apparent magnitudes, and morphological types of the galaxies were taken from the updated
HyperLeda database\footnote{\tt http://leda.univ-lyon1.fr/} and the
NED database,\footnote{\tt http://ned.ipac.caltech.edu/}
supplemented by the velocity measurements from the SDSS,
6dF, HIPASS, and ALFALFA surveys. Clustering was analyzed for
the galaxies with the line-of-sight velocities relative to the Local Group
of  \mbox{$V_{\rm LG}<4000$~km\,s$^{-1}$}, and for the galaxies within the galactic latitude
of $|b|>10\deg$. A particular benefit of the clustering algorithm proposed by the authors is 
taking into account the individual characteristics of galaxies, especially, the indicator  of the
galaxy mass---the near-infrared $K$-band luminosity. By linking the galaxies into systems pairwise, 
the authors suppose that every virtual pair has to satisfy the negative total energy condition 
and the condition of embedding its components within the ``zero velocity sphere'', i.e. the galaxies 
of the pairs do not recede from each other as a result of the Hubble expansion.
The clustering algorithm involves an iterative analysis of all galaxies of the
original sample for their further integration into the groups or clusters through 
identifying the bound pairs of galaxies having common members.
The details of the algorithm are given in the papers mentioned above. One of
the by-products of the whole procedure is a list of pairwise
isolation indices calculated between any two galaxies of the sample. The
isolation index ($II$) of two galaxies is the number characterizing the mutual dynamic
influence of components on each other. In the case of an unbound pair, $\log II$ is
positive and equal to the logarithm of the factor by which it is necessary to increase 
the mass of one of the components so that the pair would meet the given association
criteria. And conversely, in the case of a bound pair, $\log II$ is negative and equal 
to the logarithm of the factor by which the mass has to be reduced to become unbound. 
The same value of the isolation index may be found for a wide pair of
galaxies of comparable luminosities and for a tight pair consisting
of a giant galaxy and a close faint companion.


The authors of this approach have kindly provided us with the
information about the mutual isolation indices for all galaxies of the
Local Supercluster and its environs. By using this information, we
have selected for our study the lenticular galaxies having $II>2.5$ 
for the galaxies of higher and lower luminosity if compared to the considered one. 

We have made deep long-slit spectroscopic observations of 12 galaxies
from the resulting list in order to study the properties of their
stellar populations. Table~\ref{table_env} lists five the
most important neighbours of higher and lower luminosity for every
galaxy observed by us so far as well as the information about the morphology,
line-of-sight velocities, and absolute magnitudes. In the cases of the
NGC\,16 and NGC\,3098 galaxies, the conditions on the isolation 
from the potential companions of lower luminosities are not satisfied.
Both galaxies have low-luminosity neighbours with $II=1.9$; but
due to a very large luminosity difference, $\delta M_K\approx 5$ 
(the masses differ by a factor of nearly 100), we believe that the
companions cannot have a significant gravitational effect on large
galaxies, so these galaxies do not  egress from the generally
accepted selection criteria for isolated galaxies.


\begin{table*}
\centering
{\scriptsize{}

\caption{Environmental properties of galaxies.} \label{table_env}
\begin{tabular}{lcccc|lccc|lccc}
\hline\hline

\multirow{2}{*}{Galaxy} & \multicolumn{2}{c}{Type}& \multirow{2}{*}{$V_{sys}$} & \multirow{2}{*}{$M_K$} & \multicolumn{4}{c}{Upper galaxies} & \multicolumn{4}{c}{Downer galaxies} \\ 
\cline{2-3}       \cline{6-9}    \cline{10-13}    
 & HyperLEDA & NED &  &  & Galaxy & $\delta M_K$ & $\delta V$ & $II$ & Galaxy & $\delta M_K$ & $\delta V$  & $II$ \\
\hline
\hline
\multirow{3}{*}{IC0875} & \multirow{3}{*}{S0} & \multirow{3}{*}{S0} & \multirow{3}{*}{2913} & \multirow{3}{*}{-22.713} & NGC4814 &   1.034 &  278  &  31.0 & PGC046033 &  -3.949 &  -48  &   4.2 \\
                    &&&                     &                        & NGC5218 &   1.175 & -121  & 107.9 & SDSSJ1324...  &  -3.577 & -126 &  10.3 \\
                    &&&                     &                        & NGC5322 &   2.242 &  993  & 126.9 & NGC4964 &  -0.851 &  268 &   57.3 \\
                    &&&                     &                        & NGC5430 &   1.491 & -224  & 150.4 & NGC5109 &  -1.563 &  660 &   78.6 \\
                    &&&                     &                        & UGC08237 &   0.315 &  -86  & 179.2 & SDSSJ1259...     &  -4.268 &  -83 &  84.0 \\
                    
\hline
\multirow{3}{*}{IC1502} & \multirow{3}{*}{S0-a} & \multirow{3}{*}{S0\string^+} & \multirow{3}{*}{2237} & \multirow{3}{*}{-23.704} & IC0356 &   1.793 & 1131  & 158.6 & UGC12247 &  -4.960 &  298  &  46.2 \\
                    &&&                     &                        & NGC3031 &  -0.394 & 2135  &   280.1 & UGC12504 &  -4.322 & -461 &  64.6 \\
                    &&&                     &                        & NGC1184 &   0.925 & -264  &   311.8 & UGC12921 &  -2.427 & -477 &  73.5 \\
                    &&&                     &                        & IC0342 &  -0.361 & 1997  &   376.8  & UGC12160 &  -1.135 &  404 &  74.2 \\
                    &&&                     &                        & NGC6951 &   1.074 &  517  &   473.3 & UGC12069 &  -2.052 & -405 & 106.8 \\
\hline
	\multirow{3}{*}{NGC0016} & \multirow{3}{*}{E-S0} & \multirow{3}{*}{SAB0\string^-} & \multirow{3}{*}{3300} & \multirow{3}{*}{-24.511} & NGC7817 &  -0.190 &  745  & 214.2 & PGC000446 &  -5.443 & -140 &  1.9 \\
                    &&&                     &                        & NGC0253 &  -0.100 & 3024  &   1029.7 & UGC12873  &  -5.043 & -212 &   22.8 \\
                    &&edge-on&                     &                 & NGC3031 &  -1.201 & 3198  &   1212.8 & PGC087206 &  -5.736 & -513 &  191.4 \\
                    &&&                     &                        & NGC7619 &   1.071 & -689  &   1296.4 & PGC001153 &  -6.733 & -636 &  312.1 \\
                    &&&                     &                        & NGC7331 &   0.410 & 2179  &   1339.5 & UGC00285  &  -3.498 &  872 &  459.0 \\
\hline
\multirow{3}{*}{NGC2350} & \multirow{3}{*}{S0-a} & \multirow{3}{*}{S0/a} & \multirow{3}{*}{1774} & \multirow{3}{*}{-22.725} & NGC2339 &   1.168 & -396 & 135.1 & UGC03775 &  -4.638 & -244 &  12.1 \\
                    &&&                     &                        & NGC2365 &   0.681 & -429  &   400.0 & UGC03691   &  -0.708 & -327 &  81.7 \\
                    &&&                     &                        & NGC3031 &   0.584 & 1672  &   402.6 & PGC097214  &  -2.601 & -298 & 102.4 \\
                    &&&                     &                        & NGC4472 &   2.552 &  901  &   635.7 & PGC2807004 &  -5.452 &   40 & 172.1 \\
                    &&&                     &                        & IC0342  &   0.618 & 1534  &   1054.5 & --        &  -5.371 & -154 & 470.6 \\
\hline
\multirow{3}{*}{NGC3098} & \multirow{3}{*}{S0-a} & \multirow{3}{*}{S0} & \multirow{3}{*}{1305} & \multirow{3}{*}{-22.170} & NGC3190 &   1.456 &  117 &   4.1 & PGC2806869 &  -4.981 &   58 &   1.9 \\
                    &&&                     &                        & NGC3245 &   1.216 &   29  &  27.1 & IC2520    &  -1.545 &  151 &   24.1 \\
                    &&edge-on&                     &                 & NGC3227 &   0.955 &  271  &  40.4 & UGC05588  &  -2.090 &   95 &   33.0 \\
                    &&&                     &                        & NGC2964 &   1.033 &   44  &  49.0 & PGC029347 &  -2.527 &   10 &   37.5 \\
                    &&&                     &                        & NGC3379 &   1.618 &  544  &  53.3 & NGC3026   &  -1.433 & -109 &   38.5 \\
\hline
\multirow{3}{*}{NGC3248} & \multirow{3}{*}{S0} & \multirow{3}{*}{S0} & \multirow{3}{*}{1356} & \multirow{3}{*}{-21.818} & NGC3190 &   1.808 &  168 &   2.5 & PGC166105 &  -5.029 &  252 &   13.2 \\
                    &&&                     &                        & NGC3301 &   0.782 &  116  &    6.3 & PGC2806870 &  -5.424 & -115 &  16.0 \\
                    &&&                     &                        & NGC3245 &   1.568 &   80  &   15.9 & PGC030270  &  -3.603 &  -78 &  24.1 \\
                    &&&                     &                        & NGC3227 &   1.307 &  321  &   21.4 & UGC05588   &  -1.738 &  145 &  35.6 \\
                    &&&                     &                        & NGC3379 &   1.970 &  594  &   38.5 & PGC031387  &  -2.524 &   47 &  36.9 \\
\hline
\multirow{3}{*}{NGC6615} & \multirow{3}{*}{S0-a} & \multirow{3}{*}{SB0\string^+?} & \multirow{3}{*}{2868} & \multirow{3}{*}{-23.779} & NGC6574 &   0.520 &  396 &  21.7 & UGC11214 &  -1.687 &   34 &   2.5 \\
                    &&&                     &                        & UGC11057&  -0.299 & -161  &   92.4 & PGC061685 &  -2.895 & -198 &   2.9 \\
                    &&&                     &                        & NGC6548 &   0.244 &  498  &  110.9 & PGC061658 &  -1.804 & -319 &  13.5 \\
                    &&&                     &                        & NGC6501 &   0.432 & -224  &  123.7 & UGC11168  &  -0.981 &  381 &  22.4 \\
                    &&&                     &                        & NGC6587 &   0.390 & -429  &  138.0 & PGC061621 &  -0.776 & -274 &  29.4 \\
\hline
\multirow{3}{*}{NGC6654} & \multirow{3}{*}{S0-a} & \multirow{3}{*}{(R')SB0/a(s)} & \multirow{3}{*}{2204} & \multirow{3}{*}{-23.830} & NGC6643 &  -0.203 &  459 &  21.1 & PGC062387 &  -3.539 &  101 &  18.4 \\
                    &&&                     &                        & NGC6340 &  -0.651 &  757  &  189.7 & NGC6654A &  -3.756 &  380 &  25.6 \\
                    &&&                     &                        & NGC6951 &   0.948 &  484  &  225.3 & UGC10892 &  -4.568 &   26 &  33.3 \\
                    &&&                     &                        & NGC3031 &  -0.520 & 2102  &  273.3 & PGC062173&  -3.925 &  534 &  51.3 \\
                    &&&                     &                        & NGC6911 &   0.517 & -576  &  470.4 & UGC11295 &  -4.629 & -427 &  57.8 \\
\hline
\multirow{3}{*}{NGC6798} & \multirow{3}{*}{S0} & \multirow{3}{*}{S0} & \multirow{3}{*}{2741} & \multirow{3}{*}{-23.520} & NGC6764 &  -0.001 &   41 &  31.1 & UGC11457 &  -4.740 &  -10 &  6.1 \\
                    &&&                     &                        & NGC6824 &   1.934 & -1091 & 232.1 & NGC6757  &  -0.813 &   75 &  36.9 \\
                    &&&                     &                        & NGC6703 &   1.045 &   90  & 344.4 & UGC11502 &   0.047 & -335 & 112.3 \\
                    &&&                     &                        & NGC6829 &   0.376 & -853  & 505.2 & PGC063313&   0.604 & -1241& 193.8 \\
                    &&&                     &                        & NGC6946 &   0.083 & 2389  & 577.0 & NGC6796  &  -0.424 &  263 & 303.2 \\
\hline
\multirow{3}{*}{NGC7351} & \multirow{3}{*}{S0} & \multirow{3}{*}{SAB0\string^0?} & \multirow{3}{*}{1077} & \multirow{3}{*}{-20.923} & NGC0253 &   3.488 &  801 &  137.0 & PGC069415 &  -1.826 &   61 &   2.5 \\
                    &&&                     &                        & NGC7727 &   3.637 & -881  &  899.2 & PGC069224  &  -1.777 &   -7 &    3.4 \\
                    &&&                     &                        & NGC3031 &   2.387 &  975  &  1006.1 & PGC1028063&  -3.405 &   23 &   11.9 \\
                    &&&                     &                        & IC1459  &   4.208 & -599  &  1042.6 & PGC982181 &  -5.031 &  -37 &  169.8 \\
                    &&&                     &                        & NGC7507 &   3.518 & -558  &  1163.6 & PGC069293 &   0.579 & -809 &  626.0 \\
\hline
\multirow{3}{*}{UGC04551} & \multirow{3}{*}{S0} & \multirow{3}{*}{S0?} & \multirow{3}{*}{1794} & \multirow{3}{*}{-22.633} & NGC2712 &   0.334 &  -50 &  66.2 & UGC04659 &  -2.959 &   17 &  26.7 \\
                    &&&                     &                        & NGC2768 &   1.827 &  328  &   69.3 & PGC023834&  -3.120 &  -10 &   27.5 \\
                    &&&                     &                        & NGC2639 &   2.259 & -1458 &   71.1 & UGC04543 &  -1.791 & -189 &   58.5 \\
                    &&&                     &                        & NGC3031 &   0.677 & 1691  &  127.7 & UGC04648 &  -3.851 & -138 &   93.2 \\
                    &&&                     &                        & NGC2681 &   0.009 & 1048  &  159.4 & UGC04922 &  -1.653 & -228 &  168.9 \\
\hline
\multirow{3}{*}{UGC09519} & \multirow{3}{*}{S0} & \multirow{3}{*}{S0?} & \multirow{3}{*}{1782} & \multirow{3}{*}{-21.710} & NGC4472 &   3.567 &  909 &   88.7 & NGC5727 &  -2.425 &  206 &  42.1 \\
                    &&&                     &                        & NGC5353 &   3.466 & -457  & 131.0 & PGC2080256&  -3.641 & -196 &   69.1 \\
                    &&&                     &                        & NGC5611 &   0.804 & -243  & 152.5 & PGC052694 &  -2.617 &  132 &   72.0 \\
                    &&&                     &                        & NGC5582 &   0.986 &  247  & 159.8 & UGC09597  &  -3.886 &  -30 &  123.5 \\
                    &&&                     &                        & NGC5194 &   2.284 & 1221  & 194.7 & NGC5798   &  -0.596 &  -88 &  127.6 \\
\hline
\hline
\end{tabular}
}
\end{table*}



\section{OBSERVATIONS AND DATA ANALYSIS}\label{observations:Katkov_n}
\subsection{Observations}

The observations of the sample of isolated lenticular galaxies were carried
out in the prime focus of the 6-m Russian telescope of the SAO RAS
over the period of \linebreak 2011--2012. The spectral data for
all the galaxies, except NGC\,6615 and NGC\,6654, were obtained
with the \mbox{SCORPIO-2}  focal reducer \citep{scorpio2}
in the long-slit mode with the slit width of 1 arcsec and the slit length of 6.1 arcminutes. 
For the observations we have used the VPHG\,1200@540 -- volume-phase holographic grating (the grism), 
which provides a spectral resolution of ${\rm FWHM}\approx 4$~\AA\ in the range of
3800--7300~\AA. This spectral range includes both strong absorption lines, like  Mg, Fe, \mbox{G-band,} 
and a number of ionized-gas emission lines (H$\alpha$, H$\beta$, [O\,III], [N\,II], etc.) that allows 
to explore both the kinematics, age, and chemical composition of the stellar component and the kinematics
and excitation of the ionized gas. We used the CCD detector E2V\,42-90 with the  chip  size \mbox{2k$\times$4k} 
which at the \mbox{1$\times$2} binning mode readout provides a spatial scale along the slit of $0\farcs357$ per pixel 
and reciprocal dispersion of 0.86~\AA\ per pixel. Unlike the other galaxies, NGC\,6615 and NGC\,6654 were observed 
with another instrumental configuration, namely, with the \mbox{SCORPIO} instrument \citep{scorpio1} and 
the VPHG\,2300G grism that provided a spectral resolution of 2.2~\AA\ in the range of \mbox{4800--5600~\AA}, 
and the use of the EEV\,CCD\,42-40 detector with the chip size \mbox{2k$\times$2k} gave the same scale along 
the slit with the reciprocal dispersion of 0.37~\AA\ per pixel. During all the observations the slit was aligned
along the outer-isophote major axes of the galaxies. The log of the observations (Table~\ref{obslog:Katkov_n}) gives 
the dates of observations, total exposures, average seeing conditions during the exposure
of every galaxy, and the position angle of the slit.

\subsection{Primary data reduction}

Primary data reduction was performed by using the original programs,
developed in the IDL environment, including the following steps: 
taking into account the registration system bias
by subtracting the averaged zero exposure frame from all the
images; taking into account the uneven illumination and inhomogeneities
in the CCD sensitivity through dividing the spectra by the flat-field
calibration lamp; removal of the traces of cosmic hits by
using the {\sc L.A.Cosmic} algorithm \citep{lacosmic}, which
implements the Laplacian filter for hit trace detection, and 
addition of spectra after cleaning; two-dimensional wavelength
calibration by using the spectrum of the helium-neon-argon lamp
and the further linearization of spectra with the typical
accuracy of 0.03--0.06~\AA\ depending on the grism used;
subtraction of the background spectrum of the night sky;
conversion of the instrumental fluxes into the absolute ones by using
the spectra of spectrophotometric standards. During the observations with the Moon 
high enough the night sky contribution is large and varies with time.   
In such cases we performed the night sky subtraction before the frame
additions to fit the best parameters of the sky background for each
spectrum  individually. Among the final results of the primary
data reduction there were not only the reduced spectra of the objects but also
the frames of the errors that were calculated assuming the
Poisson photon statistics and readout noise and pushed through all 
reduction steps together with the data frames.

In addition to the spectra of objects and spectrophotometric
standards, the spectra of dawn or twilight sky were also
analyzed in the process of data reduction, which, in the essence,
represent the solar spectrum dispersed by the Earth atmosphere which
are convoluted with the instrumental profile of the spectrograph in every
point of the slit. Therefore, the analysis of the spectra of the dawn or twilight sky 
allows to determine the behaviour of the instrumental profile of the spectrograph along
the slit and along the direction of dispersion. The former is important to subtract correctly
the contribution of the night sky, and the latter allows to accurately determine the 
kinematical parameters. The spectra of the dawn sky were reduced in the same manner as the 
spectra of the galaxies though without the sky background subtraction. The
reconstruction of the instrumental profile and its use when analyzing the galaxies are  
described in detail in the following Section 3.3.

\begin{table}
{\small{}
\caption{The long-slit observation log.}
\medskip
\begin{tabular}{lrccr}
\hline
\multirow{2}{*}{Name} & \multicolumn{1}{c}{\multirow{2}{*}{Date}} & Exposure, & Seeing, & \multicolumn{1}{c}{PA,} \\
      &      &  s      & arcsec & \multicolumn{1}{c}{deg}\\
\hline
IC\,875   &  Apr 23,   2012 & 2700 & 2.5 & $-$30\\
IC\,1502  &  Nov 19,  2011 & 2700 & 2.5 & 52\\
NGC\,16   &  Nov 20,   2011 & 1800 & 2.0 & 16\\
NGC\,2350 &  Dec 13,  2012 & 6000 & 1.6 & $-$73\\
NGC\,3098 &  Apr 18,   2012 & 5400 & 1.2 & $-$90\\
NGC\,3248 &  Apr 22,   2012 & 2700 & 3.0 & $-$45\\
NGC\,6615 &  Sep 19, 2012 & 7200 & 1.0 & $-$15\\
NGC\,6654 &  Sep 20, 2012 & 6600 & 1.3 & 0\\
NGC\,6798 &  Nov 20,   2011 & 5400 & 2.5 & $-$30\\
NGC\,7351 &  Nov 19,  2011 & 3600 & 2.0 & 0\\
UGC\,4551 &  Dec 12, 2012 & 8400 & 2.0 & $-$67\\
UGC\,9519 &  Apr 24,  2012 & 4500 & 2.0 & $-$105\\
\hline
\end{tabular}
\label{obslog:Katkov_n}
}
\end{table}


\subsection{Subtraction of night sky contribution}

When analyzing spectral data for objects with low surface brightness, special
attention should be paid to the careful subtraction of the
contribution of the night-sky background, underestimation of which may lead
to systematic errors in the derived parameters of stellar populations \citep{skysubtraction}.
In this paper we determine the properties of stellar populations
in the galactic structural components, including the large-scale stellar
discs of low surface brightness. Therefore, we believe it is
necessary to describe in detail the refined procedures of the night-sky
spectrum subtraction.

We have earlier proposed a refined technique for subtracting
the spectrum of the night sky for the long-slit spectral data of
the low surface brightness objects in case of strong variations
of the instrumental profile along the slit \citep{skysubtraction}. 
In that approach the sky model is constructed by interpolating  
the spectrum of the sky from the edges of the slit to the central part 
containing the spectrum of the object, by applying the
deconvolution procedure to the reference spectrum. The spectrum of the
twilight sky, which carries the information on the variations of the
instrumental profile, is supposed to be used as a reference
spectrum. Unfortunately, during the real observations the spectra of
the dawn/twilight sky were not routinely observed due to a sudden
worsening of weather conditions during the nights. The standard method which 
does not require the use of the reference spectrum and provides
zero approximation for the profile variations  along the slit,
consists in the approximation of the sky spectrum  by the
low-power polynomials ($n=2--4$) in each column of the spectrum image
parallel to the slit, and its recomputation for the area of the object. 
This method works well for the spectra of non-extended objects. However, this
straightforward approach is not always suitable for the long-slit observations of
galaxies.

Therefore, here we propose another way to build the 2D spectrum of the
night sky, which is based on the extrapolation procedure in the
frequency domain. The spectrum of the night sky at a given
position on the slit $y$--$S(\lambda,y)$ can be written as a
convolution of  the ``intrinsic'' spectrum of the night sky
$S_0(\lambda)$ with the instrumental profile
$\mathcal{L}(\lambda,y)$:
\begin{equation}
S(\lambda,y)=S_0(\lambda)\,\mathcal{L}(\lambda,y).
\label{formula_convolution:Katkov_n}
\end{equation}
In the frequency domain the convolution procedure becomes a
multiplication. Hence, if we  perform  a one-dimensional Fourier transform
along the direction of dispersion, then
\begin{equation}
{\rm FFT}[ S(\lambda,y) ] = {\rm FFT}[S_0(\lambda)]\,{\rm
FFT}[\mathcal{L}(\lambda,y)].
\end{equation}
As the analysis of the spectra of the dawn and twilight sky has
shown, the shape of the instrumental profile of the
SCORPIO/SCORPIO-2 spectrographs varies rather monotonously along
the slit, therefore its Fourier transform  ${\rm
FFT}[\mathcal{L}(\lambda,y)]$ varies along the slit 
monotonously as well. The first factor in the Fourier transform of the
night sky spectrum is a constant function, hence in general the
Fourier transform of the night sky along the slit varies
monotonously. Using this fact and applying the standard polynomial
extrapolation procedure  to the spectrum of the night sky in the
frequency domain, i.e., to its Fourier transform, and performing
the subsequent inverse Fourier transform, we can construct the
model of the spectrum of the night sky. The Fourier transform of
the  spectral image returns complex values, this is why the
extrapolation should be performed separately for the real and
imaginary parts of the Fourier transform. In this 
model of the night sky, the reference spectrum of the
dawn/twilight sky  is not required. At the same time the quality
of the model is comparable to the method based on the
deconvolution procedure~\citep{skysubtraction}.

\subsection{Data analysis}
\label{lsfmark:Katkov_n}

Before undertaking the analysis of the spectra of galaxies, we analyzed the
spectrum of the dawn sky observed the same observational run, in order to estimate 
the variations of the parameters of the spectrograph instrumental profile
(LSF---Line Spread Function) along and across the direction of
dispersion which is required to determine properly the internal
kinematics of stars and ionized gas in the galaxies. To do this, we have
split the frame of the twilight sky spectrum into many areas:
32~intervals along the slit and 7~segments along the dispersion. In
every section the twilight sky spectra were co-added to achieve the typical
signal-to-noise ratio \mbox{$S/N=100$} per pixel. Then, we fitted the
spectra from every area by the high-resolution solar spectrum,
taken from the ELODIE\,3.1  stellar spectral library~\citep{elodie3.1}, 
by applying the {\sc ppxf} procedure of the per-pixel approximation of 
spectra~\citep{ppxf}. During fitting the twilight sky spectrum, the 
instrumental profile is parametrized by the Gauss--Hermite series of
orthogonal functions~\citep{gausshermite}. As a result of
approximation, we derive the instrumental profile (LSF)
parameters for different spectral ranges and along the slit. The mean
characteristic widths of the instrumental profile (in terms of
velocity dispersion) for the observational mode with the
VPHG\,2300G and  VPHG\,1200@540 grisms are \mbox{$\sigma_{\rm
instr}=65$~km\,s$^{-1}$} and \mbox{$\sigma_{\rm
instr}=90$~km\,s$^{-1}$} respectively.

The further analysis consisted of fitting the
observed absorption-line spectra of galaxies by the high-resolution
models of stellar populations. To do this, we used the {\sc NBursts} software
package~\citep{nbursts_a,nbursts_b}, which represents an extension of the 
{\sc ppxf} per-pixel spectrum approximation method~\citep{ppxf}. 
The approach applied allows to derive the information about the stellar component 
from all the available spectral elements in total, in contrast to the
analysis of the Lick indices of individual lines~\citep{licksystem1,licksystem2}---the indicators
of the properties of stellar populations. The approach of
per-pixel approximation of the spectra of galaxies allows to
exclude from the analysis the narrow  spectral ranges around strong
emission lines of ionized gas, it allows to avoid their systematic
effect on the estimates of the stellar population parameters
which is impossible when analyzing the Lick index H$\beta$. \citet{nbursts_a,nbursts_b} have shown that the per-pixel spectrum 
approximation method provides also a higher accuracy, by a factor of 1.5--2, for
the stellar population parameters derived, in comparison to the approach of the Lick indices.

\renewcommand{\baselinestretch}{0.81}

\begin{table*}
\caption{The mean stellar population parameters of the bulges, discs, and
lenses/rings} \label{table_stpop:Katkov_n}
\medskip
\begin{tabular}{l|c|c|c|c|c|c}

\hline
\multirow{2}{*}{Galaxy} & Averaging range,& Number & $T$, & [Z/H], & [Mg/Fe],  & $\sigma$,  \\
 & arcsec & of measurements & Gyr&dex&dex& km\,s$^{-1}$\\
\hline

\multicolumn{7}{c}{Bulge}\\
\hline
IC\,875 & $4$--$7$ &     10 & $    4.3^{\pm 0.7}$ & $   -0.16^{\pm 0.05}$ & $    0.20^{\pm 0.04}$ & $  110^{\pm  9}$ \\
IC\,1502 & $4$--$7$ &     11 & $   17.6^{\pm 0.9}$ & $   -0.04^{\pm 0.06}$ & $    0.3^{\pm 0.1}$ & $  168^{\pm 16}$ \\
NGC\,16 & $2$--$5$ &     10 & $    5.4^{\pm 0.8}$ & $   -0.04^{\pm 0.05}$ & $    0.19^{\pm 0.04}$ & $  172^{\pm  6}$ \\
NGC\,2350 & $4$--$7$ &      8 & $    1.6^{\pm 0.3}$ & $   -0.13^{\pm 0.08}$ & -- & $  103^{\pm 15}$ \\
NGC\,3098 & $4$--$7$ &      8 & $    5.4^{\pm 0.4}$ & $   -0.10^{\pm 0.02}$ & $    0.00^{\pm 0.02}$ & $   73^{\pm  6}$ \\
NGC\,3248 & $4$--$7$ &     10 & $    4.8^{\pm 0.6}$ & $   -0.11^{\pm 0.05}$ & $    0.00^{\pm 0.05}$ & $   77^{\pm  5}$ \\
NGC\,6615 & $4$--$7$ &      8 & $   10.8^{\pm 1.5}$ & $   -0.26^{\pm 0.05}$ & $    0.24^{\pm 0.03}$ & $  129^{\pm  5}$ \\
NGC\,6654 & $2$--$5$ &      9 & $   12.2^{\pm 1.4}$ & $   -0.19^{\pm 0.07}$ & $    0.23^{\pm 0.04}$ & $  158^{\pm  5}$ \\
NGC\,6798 & $4$--$7$ &      6 & $    8.0^{\pm 1.9}$ & $   -0.20^{\pm 0.05}$ & $    0.13^{\pm 0.04}$ & $  115^{\pm  7}$ \\
NGC\,7351 & $4$--$7$ &     10 & $    2.2^{\pm 0.5}$ & $   -0.37^{\pm 0.08}$ & $   -0.03^{\pm 0.06}$ & $   29^{\pm 11}$ \\
UGC\,4551 & $4$--$7$ &      8 & $    10.0^{\pm 1.9}$ & $   -0.28^{\pm 0.08}$ & $    0.15^{\pm 0.03}$ & $  158^{\pm 11}$ \\
UGC\,9519 & $4$--$7$ &      9 & $    2.5^{\pm 0.1}$ & $   -0.12^{\pm 0.06}$ & $    0.04^{\pm 0.03}$ & $   76^{\pm  3}$ \\
\hline
\multicolumn{7}{c}{Disc}\\
\hline
IC\,875 & $13$--$45$ &     16 & $    2.9^{\pm 0.9}$ & $   -0.32^{\pm 0.18}$ & $    0.26^{\pm 0.07}$ & $  134^{\pm 27}$ \\
IC\,1502 & $7$--$25$ &     11 & $   16.7^{\pm 1.6}$ & $   -0.13^{\pm 0.10}$ & $    0.42^{\pm 0.01}$ & $  130^{\pm 25}$ \\
NGC\,16 & $6$--$30$ &     18 & $    1.6^{\pm 1.2}$ & $   -0.19^{\pm 0.15}$ & $    0.16^{\pm 0.02}$ & $  127^{\pm 18}$ \\
NGC\,2350 & $10$--$40$ &     15 & $    1.3^{\pm 0.2}$ & $   -0.00^{\pm 0.07}$ & $    0.06^{\pm 0.07}$ & $   86^{\pm 14}$ \\
NGC\,3098 & $25$--$60$ &     18 & $    5.1^{\pm 1.5}$ & $   -0.22^{\pm 0.06}$ & $    0.08^{\pm 0.02}$ & $   57^{\pm 26}$ \\
NGC\,3248 & $10$--$39$ &     31 & $    3.9^{\pm 1.4}$ & $   -0.21^{\pm 0.09}$ & $   -0.04^{\pm 0.03}$ & $   65^{\pm 17}$ \\
NGC\,6615 & $40$--$60$ &      0 &      -- &      -- &      -- &      -- \\
NGC\,6654 & $35$--$60$ &      3 & $    5.8^{\pm 0.6}$ & $   -0.06^{\pm 0.14}$ & $    0.40^{\pm 0.20}$ & $   44^{\pm  5}$ \\
NGC\,6798 & $8$--$55$ &     18 & $    7.3^{\pm 4.3}$ & $   -0.27^{\pm 0.15}$ & $    0.11^{\pm 0.12}$ & $  119^{\pm 17}$ \\
NGC\,7351 & $17$--$45$ &      7 & $    4.4^{\pm 2.3}$ & $   -0.57^{\pm 0.19}$ & $   -0.02^{\pm 0.15}$ & $   72^{\pm 40}$ \\
UGC\,4551 & $38$--$80$ &      3 & $   10.9^{\pm 4.3}$ & $   -0.74^{\pm 0.39}$ & $    0.25^{\pm 0.25}$ & $  107^{\pm 20}$ \\
UGC\,9519 & $16$--$30$ &      4 & $    2.9^{\pm 0.9}$ & $   -0.32^{\pm 0.17}$ & $    0.15^{\pm 0.20}$ & $   98^{\pm 15}$ \\
\hline
\multicolumn{7}{c}{Lens/Ring}\\
\hline
IC\,875 & -- &      0 & -- & -- & -- & -- \\
IC\,1502 & -- &      0 &     -- &      -- &      -- &    -- \\
NGC\,16 & $12$--$21$ &     16 & $    3.3^{\pm 2.9}$ & $   -0.25^{\pm 0.16}$ &      -- & $  104^{\pm 16}$ \\
NGC\,2350 & $20$--$26$ &      1 & $    4.9^{\pm 0.0}$ & $   -0.33^{\pm 0.00}$ &      -- & $   97^{\pm  0}$ \\
NGC\,3098 & $15$--$20$ &     14 & $    4.8^{\pm 1.3}$ & $   -0.13^{\pm 0.05}$ & $    0.05^{\pm 0.01}$ & $   57^{\pm 12}$ \\
NGC\,3248 & -- &      0 &      -- &      -- &      -- & -- \\
NGC\,6615 & $20$--$40$ &      3 & $   12.8^{\pm 2.4}$ & $   -0.52^{\pm 0.16}$ & $    0.21^{\pm 0.06}$ & $   56^{\pm  5}$ \\
NGC\,6654 & -- &      0 &      -- &      -- &      -- & -- \\
NGC\,6798 & $14$--$22$ &     10 & $    5.2^{\pm 2.1}$ & $   -0.30^{\pm 0.14}$ & $    0.13^{\pm 0.04}$ & $   97^{\pm 15}$ \\
NGC\,7351 & -- &      0 &      -- &      -- &      -- & -- \\
UGC\,4551 & $17$--$35$ &     12 & $    3.3^{\pm 2.2}$ & $   -0.47^{\pm 0.23}$ & $    0.23^{\pm 0.03}$ & $  117^{\pm 25}$ \\
UGC\,9519 & $7$--$15$ &     22 & $    2.7^{\pm 0.5}$ & $   -0.22^{\pm 0.07}$ & $    0.05^{\pm 0.02}$ & $   77^{\pm  9}$ \\
\hline
 \end{tabular}
 \end{table*}

\renewcommand{\baselinestretch}{1.00}

The core of the procedure of determining the parameters of the stellar
populations represents nonlinear minimization of the quadratic
difference  $\left(\chi^2\right)$  between the observed and model
spectra. We have used synthetic spectra of stellar populations as
model spectra. They were computed with the {\sc Pegase.HR}~\citep{pegasehr}  
evolutionary code based on the  ELODIE\,3.1 high-resolution stellar spectral
library~\citep{elodie3.1} for the simple star formation history in the form 
of one brief burst (SSP, Simple Stellar Population). The parameters of the SSP 
stellar population model are the age of the star formation burst $T$~(Gyr) 
and the metallicity [Z/H]~(dex), while the Salpeter initial stellar mass
function~\citep{SalpeterIMF} and solar chemical element ratios are considered 
to be fixed in the model. To measure the stellar kinematics of the galaxy, 
line-of-sight velocity distributions (LOSVDs), which are parametrized as the
Gauss--Hermite quadrature~\citep{gausshermite}, are convolved with
the model spectrum. Furthermore, the multiplicative pseudo-continuum 
is included into the model that allows to take into account the
effects of the intrinsic interstellar extinction of a galaxy as well as 
the possible errors of the flux absolute calibration both in the observational 
data and in the stellar library, serving as the base on which the stellar population 
models were synthesized. To take into account the effect of the spectrograph LSF on the 
spectrum of a galaxy, before fitting the observed spectra we convoluted the grid of 
stellar population models with the previously determined instrumental profile. 
The presence of even weak emission lines and/or the residues from the inaccurate 
subtraction of the strongest night-sky lines can bias the  stellar
population parameter estimates. Hence, to eliminate this effect,
we have masked the spectral ranges of \mbox{10--15}~\AA\ around them.
As a result of the $\chi^2$  minimization, the following set of parameters
is derived: the line-of-sight stellar velocity  $v$~(km\,s$^{-1}$), the
line-of-sight stellar velocity dispersion $\sigma$~(km\,s$^{-1}$), the estimates 
of the SSP-equivalent stellar age  $T$~(Gyr) and metallicity [Z/H]~(dex).


Our spectroscopy was carried out with a long slit, which was aligned along 
the major axis of the galaxies under consideration. Since the surface brightness 
of galactic discs decreases with the distance from the center, the spectra of the 
outer regions of the galaxies have low $S/N$ ratio. To increase it, we used the
adaptive binning procedure, by co-adding the spectra over the intervals of variable 
length along the slit, arranged so that the $S/N$ ratio is to be not less than the 
preassigned value (typically about 20--30) in each interval.

After subtracting the stellar component model from  the observed
spectrum of a galaxy, we got a pure emission-line spectrum of the
galaxy. Every line was approximated by a Gaussian to obtain the
line fluxes, velocities, and  velocity dispersions of ionized gas.
In this paper we only consider the properties of the stellar
populations, while the paper~\citep{ilg_gas} is dedicated to the
properties of the ionized gas in these galaxies.

The {\sc Pegase.HR} stellar population models were synthesized
basing on the library of spectra of stars in the solar vicinity, possessing
the solar \mbox{$\alpha$-element}-to-iron abundance ratio. The resulting models
are hence only computed for the solar abundance ratios. The methods
for constructing the stellar population models taking into account non-solar 
\mbox{$\alpha$-element} abundances are currently being
developed (see, e.g., the studies \citep{Walcher_alpha}
and \citep{Prugniel_alpha}). However, these first-trial models
are calculated for a rather rare and limited grid of parameters and
are so far inferior in quality to the models with the solar
abundance ratio. The relative abundance of \mbox{$\alpha$-elements} 
bears information on the duration of the main star formation epoch 
which has provided the bulk of stars. If the burst was very brief, shorter 
than 1~Gyr, then the stellar populations would demonstrate an excess of the
\mbox{$\alpha$-elements} with respect to iron in comparison with
the solar chemical composition. Given a long history of star
formation, as in the Sun's environs, the ratio of abundances of iron and
\mbox{$\alpha$-elements}, magnesium in particular, becomes
solar \citep{tinsley79,mat_greg_86}. To estimate the relative abundance of 
\mbox{$\alpha$-elements} of the stellar populations in the galaxies under consideration, 
in addition to the per-pixel approximation method we have also undertaken 
classical Lick-index approach. For the spectra of the galaxies observed we have 
calculated the Lick H$\beta$, Mgb, Fe\,5270, and Fe\,5335 indices, the definition of which
were taken from \citep{licksystem1,licksystem2}.
By comparing the obtained indices to the model values,
calculated in the framework of stellar population synthesis
models for different values of magnesium-to-iron ratios \citep{Thomasstpop}, 
we have derived the Mg/Fe ratio values.

\begin{figure*}
\centerline{
  \includegraphics[width=0.36\textwidth]{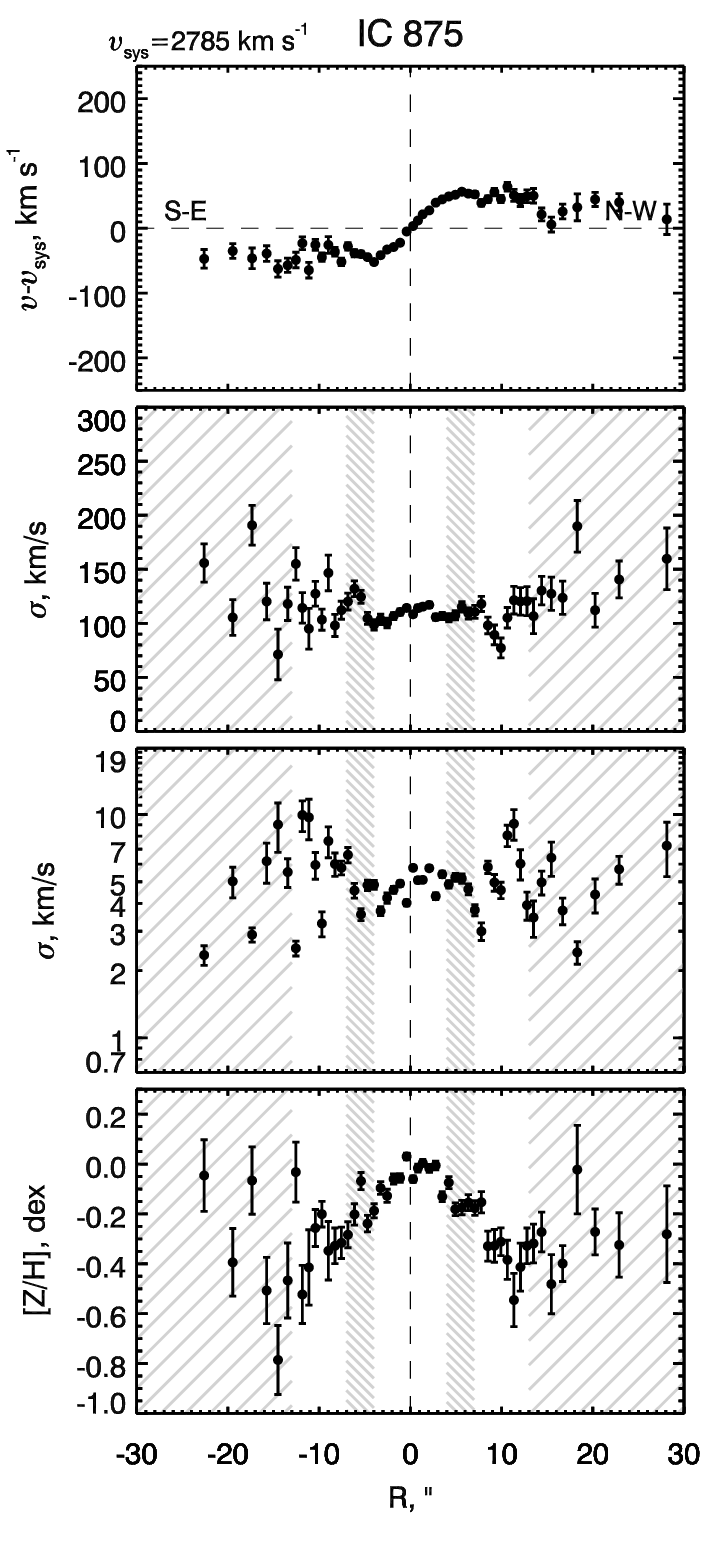}
  \hspace{-10mm}
  \includegraphics[width=0.36\textwidth]{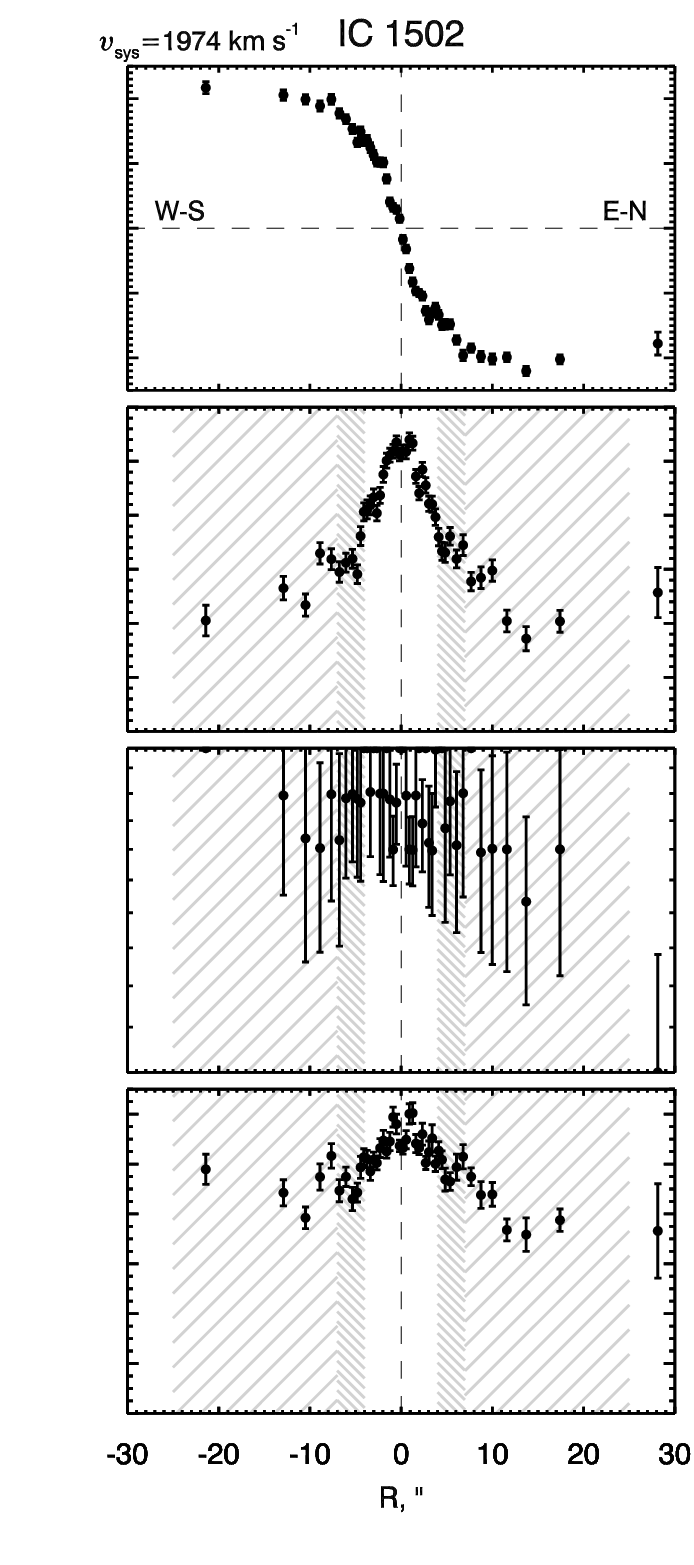}
  \hspace{-10mm}
  \includegraphics[width=0.36\textwidth]{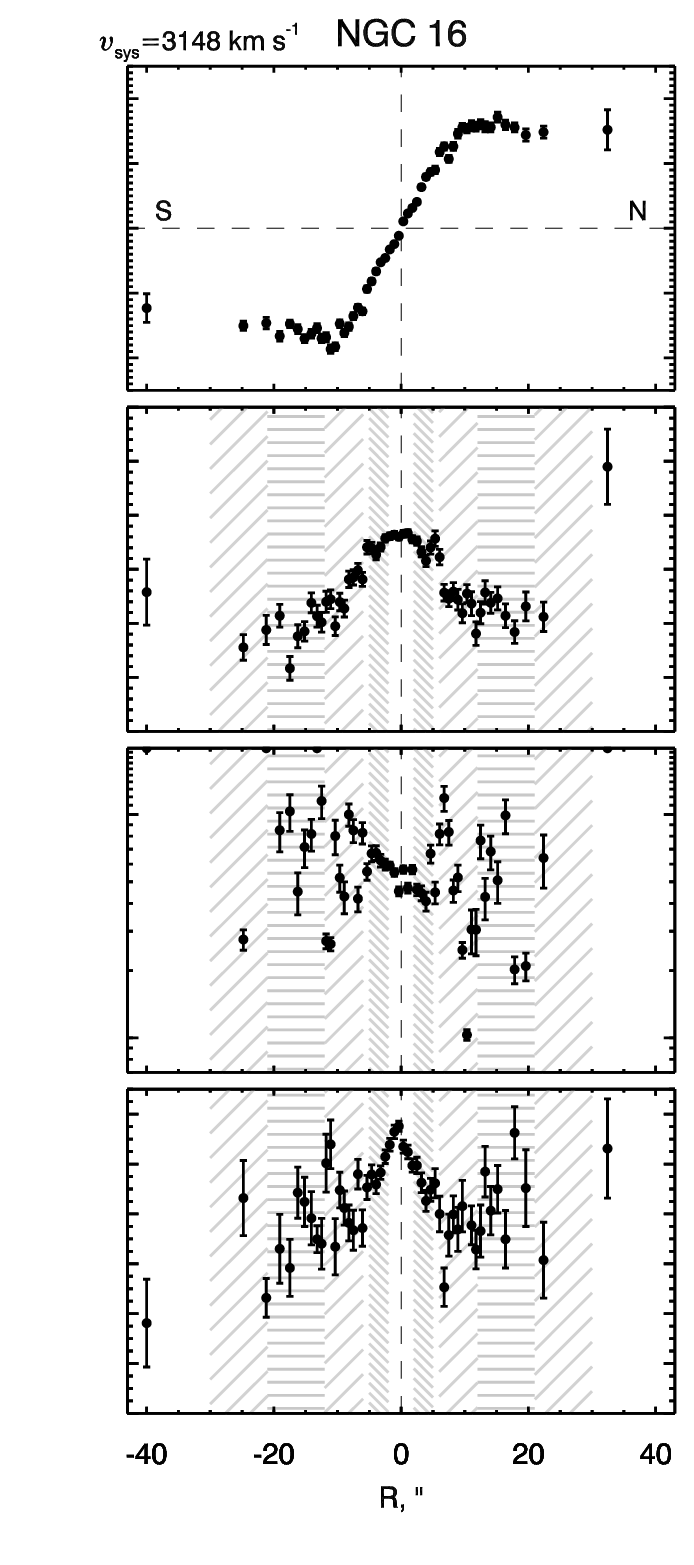}
} \caption{The results of analysis of the long-slit spectra of the
galaxies under consideration. Each column corresponds to one galaxy. The
panels with radial profiles of the line-of-sight velocity of stars
after subtracting the systematic velocity, stellar velocity dispersions,
ages, and metallicities are plotted from top to bottom. The distance
from the center of a galaxy in arcseconds is plotted along the
$x$-axis. The gray slashes mark the areas of the averaging of the
bulge parameters~(\textbackslash) and the exponential disc
parameters~(/), the horizontal lines~(---) are the regions of the
lens or the ring dominance, if present. The systematic velocity corresponds
to the measured line-of-sight velocity of the galaxy's center, i.e., the
brightest part, except for IC\,875 where this is the center of
symmetry of the central part of the velocity profile.}
\label{kin_stpop_profiles:Katkov_n}
\end{figure*}

\begin{figure*}
\centerline{
  \includegraphics[width=0.36\textwidth]{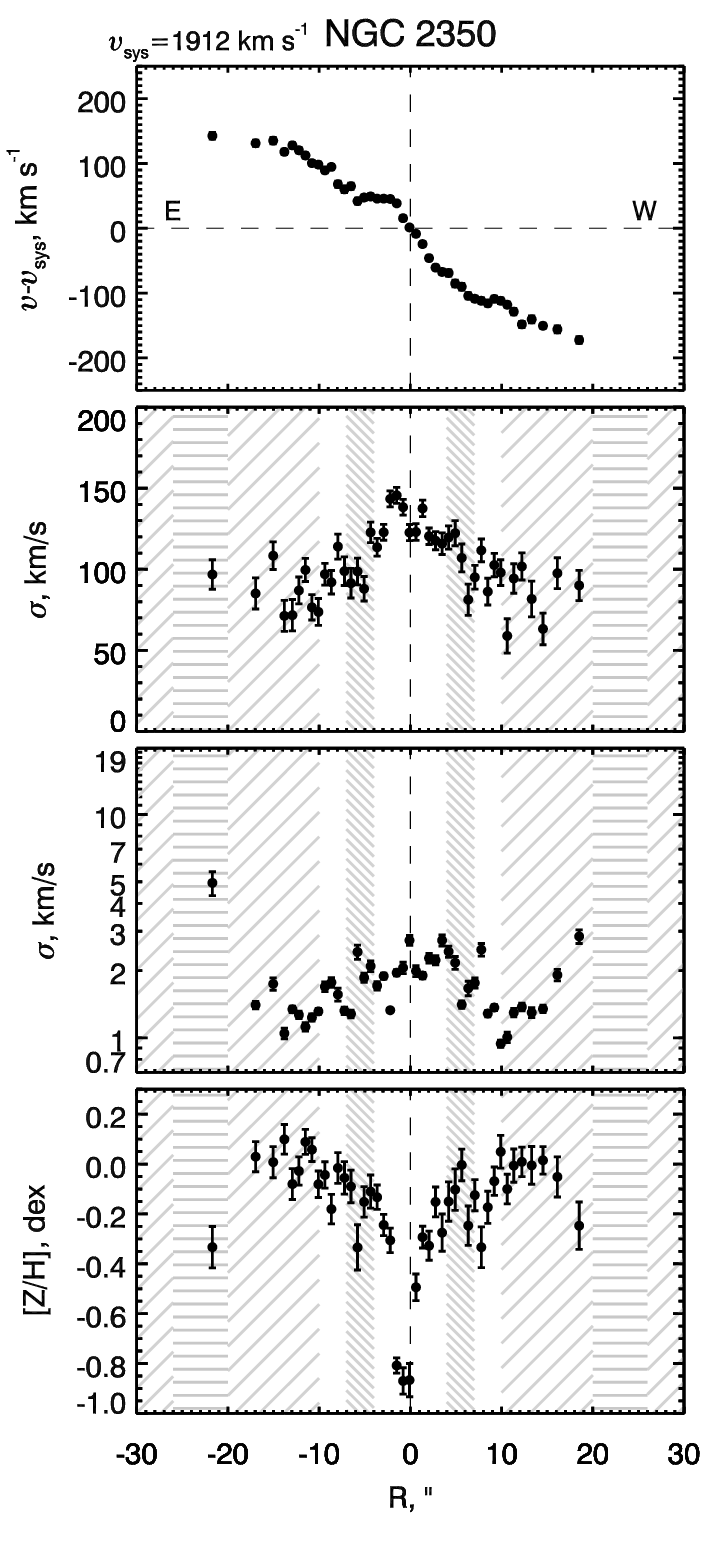}
  \hspace{-10mm}
  \includegraphics[width=0.36\textwidth]{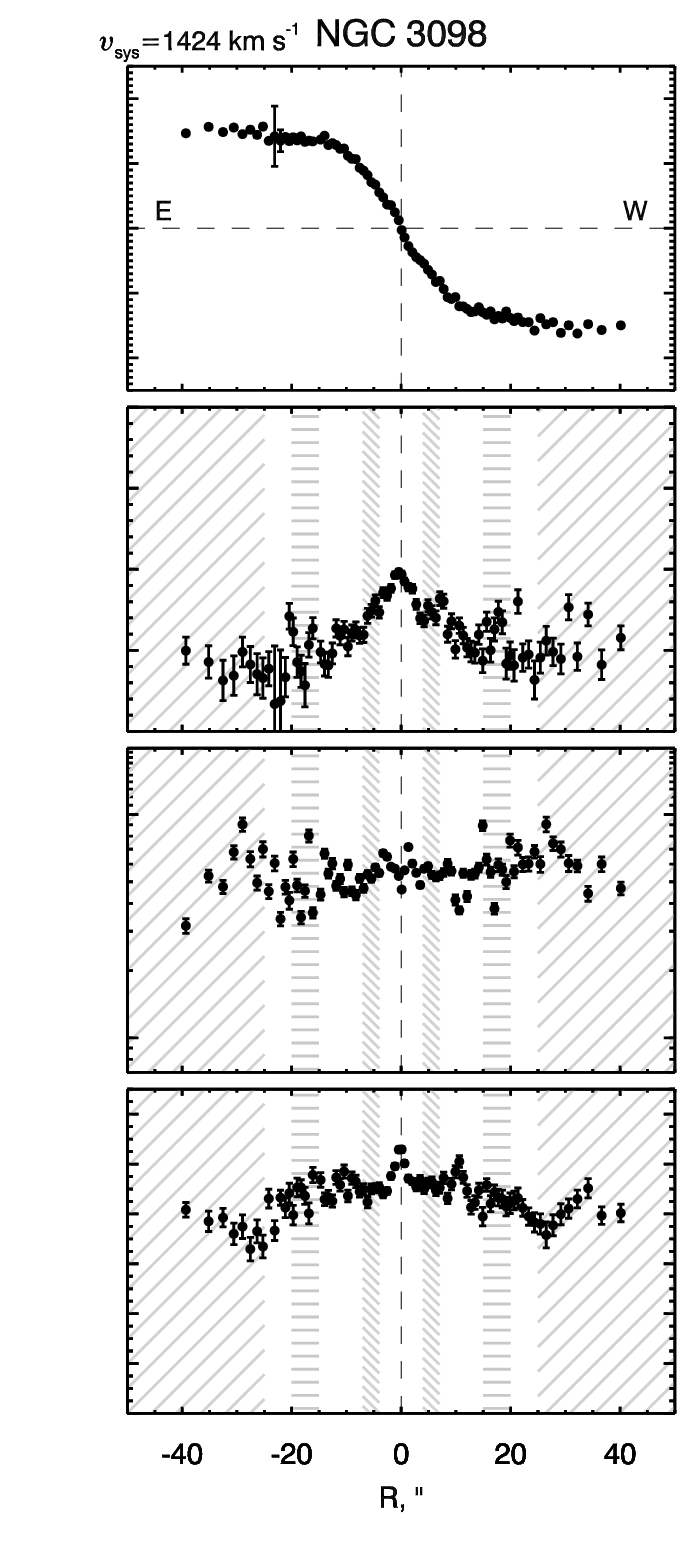}
  \hspace{-10mm}
  \includegraphics[width=0.36\textwidth]{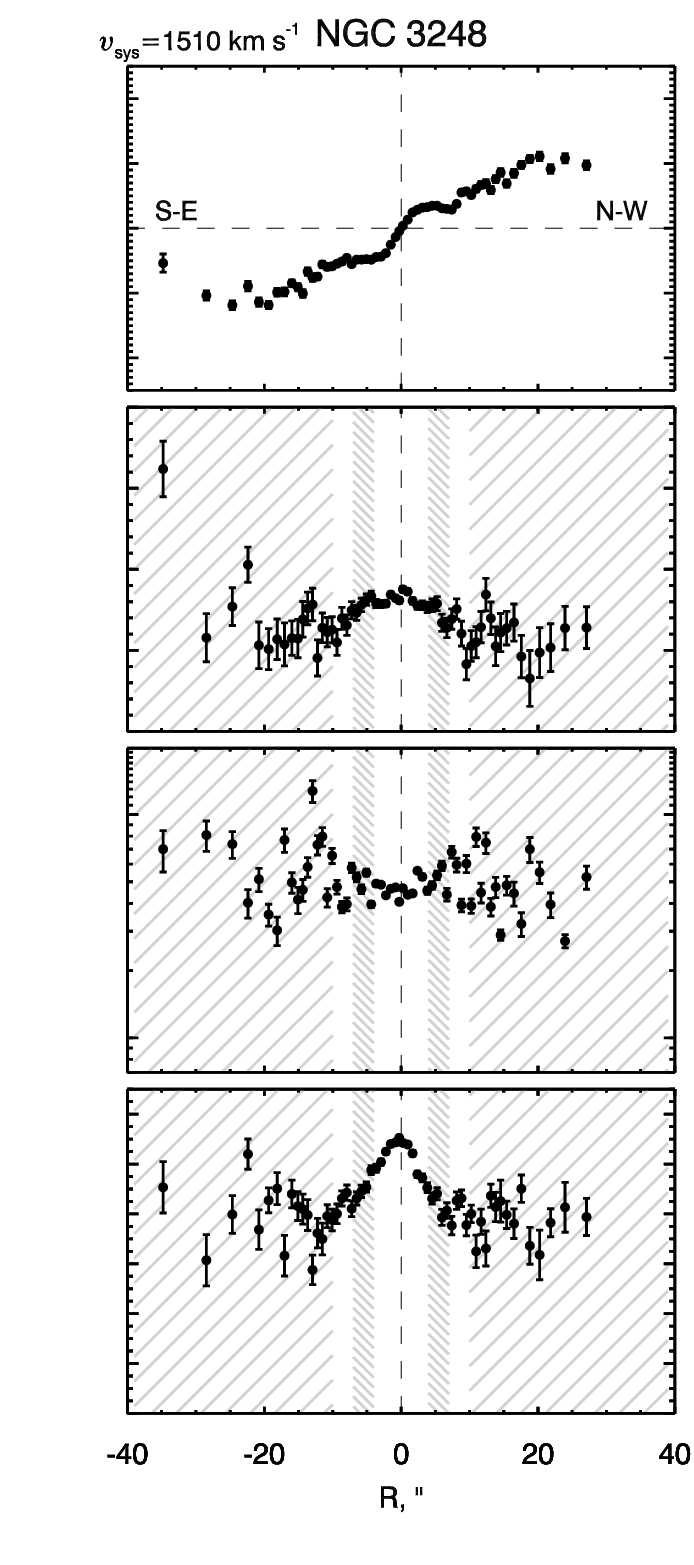}}
\addtocounter{figure}{-1}
 \caption{(Cont.)}
\end{figure*}

\begin{figure*}
\centerline{
  \includegraphics[width=0.36\textwidth]{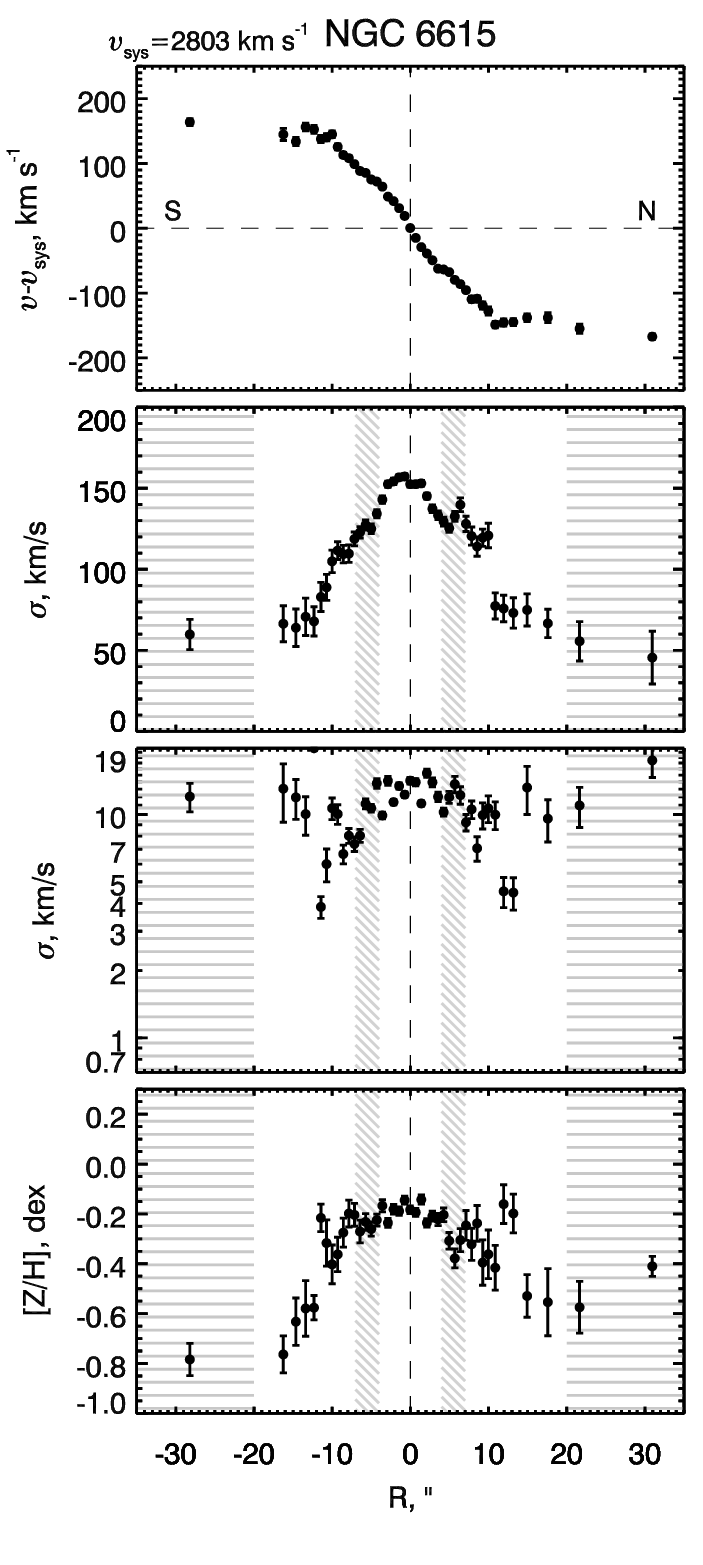}
  \hspace{-10mm}
  \includegraphics[width=0.36\textwidth]{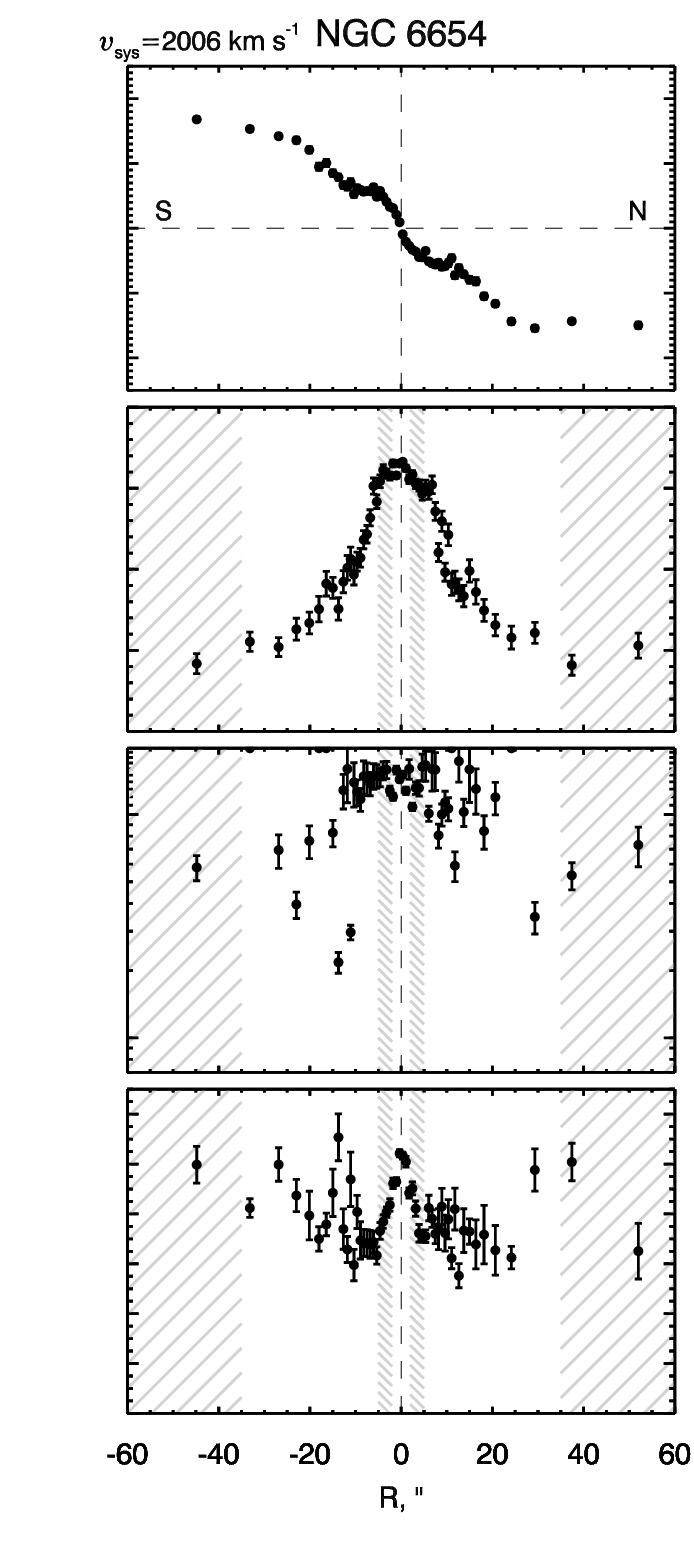}
  \hspace{-10mm}
  \includegraphics[width=0.36\textwidth]{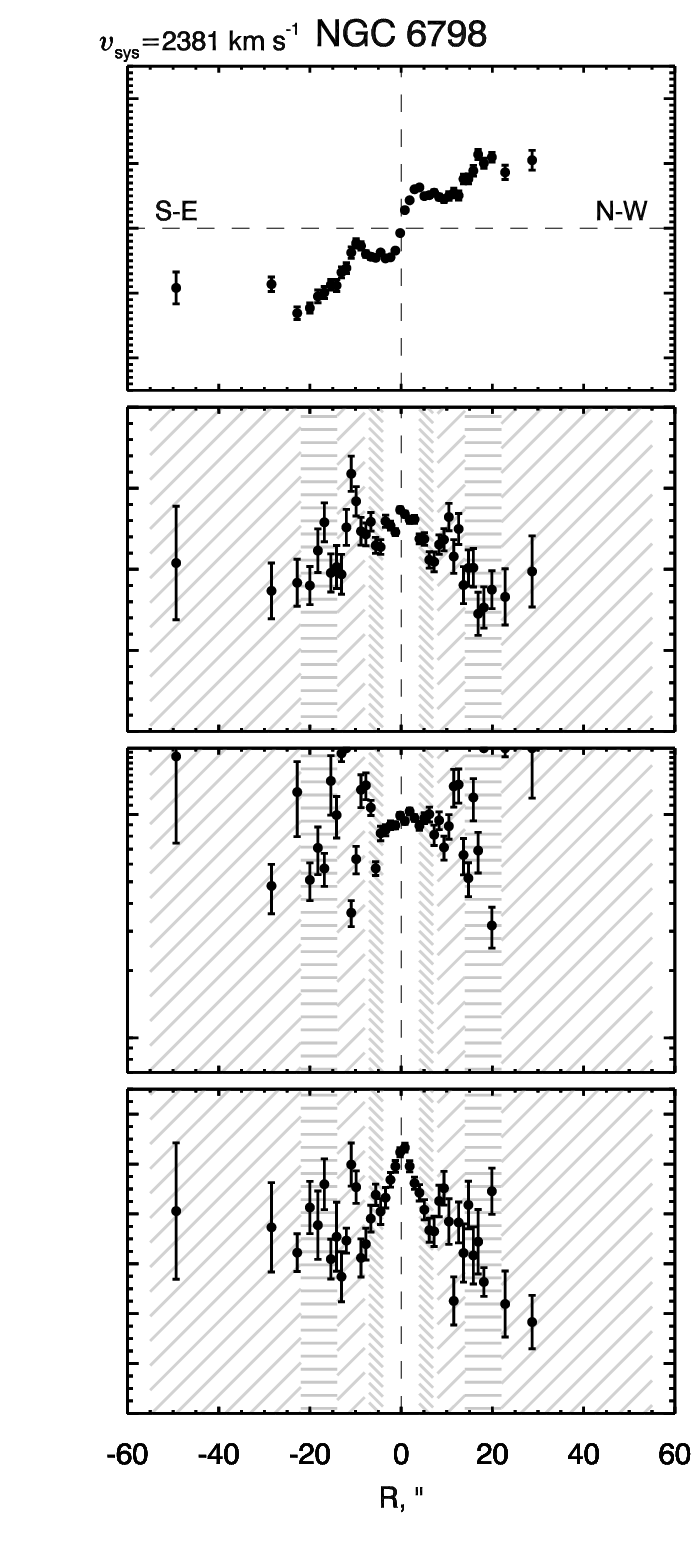}
} \addtocounter{figure}{-1}
 \caption{(Cont.)}
\end{figure*}

\begin{figure*}
\centerline{
  \includegraphics[width=0.36\textwidth]{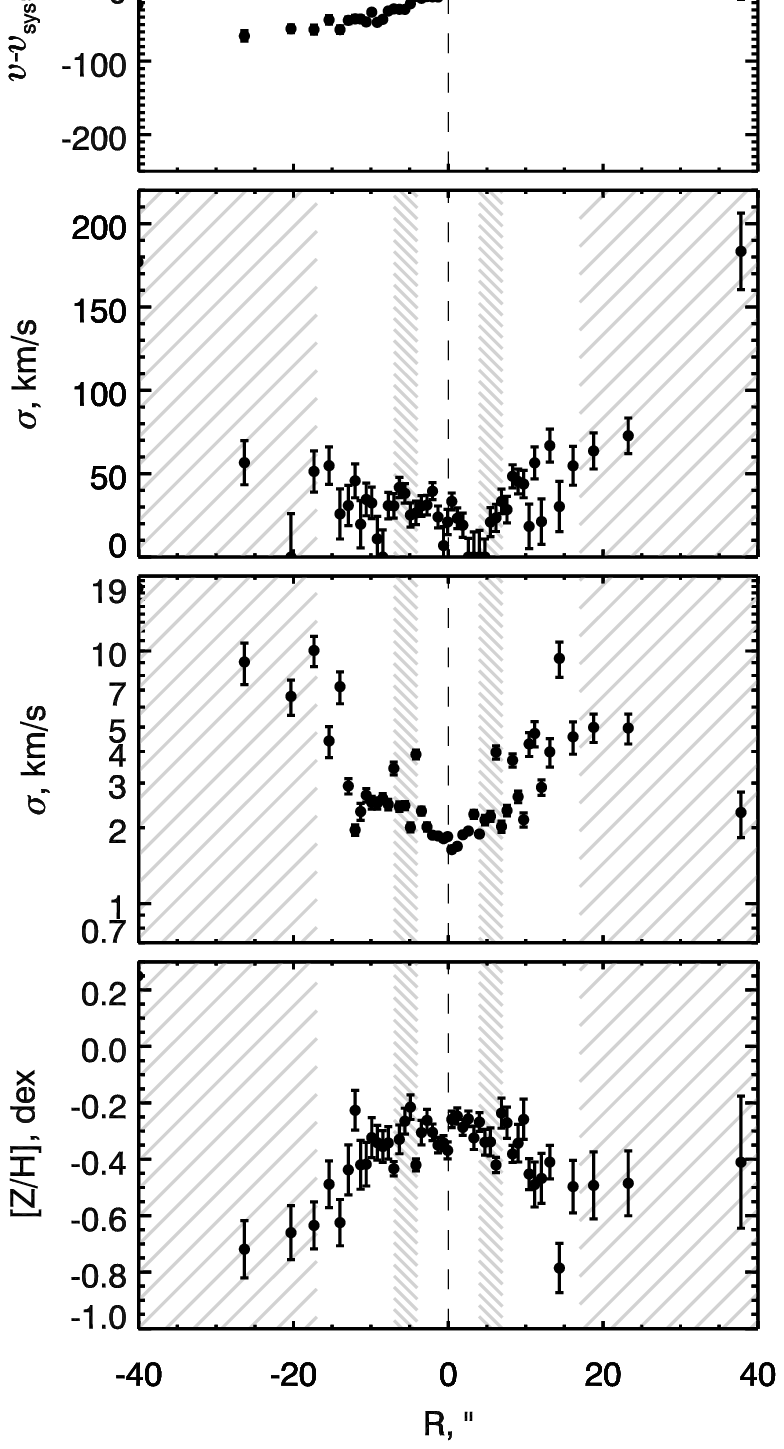}
  \hspace{-10mm}
  \includegraphics[width=0.36\textwidth]{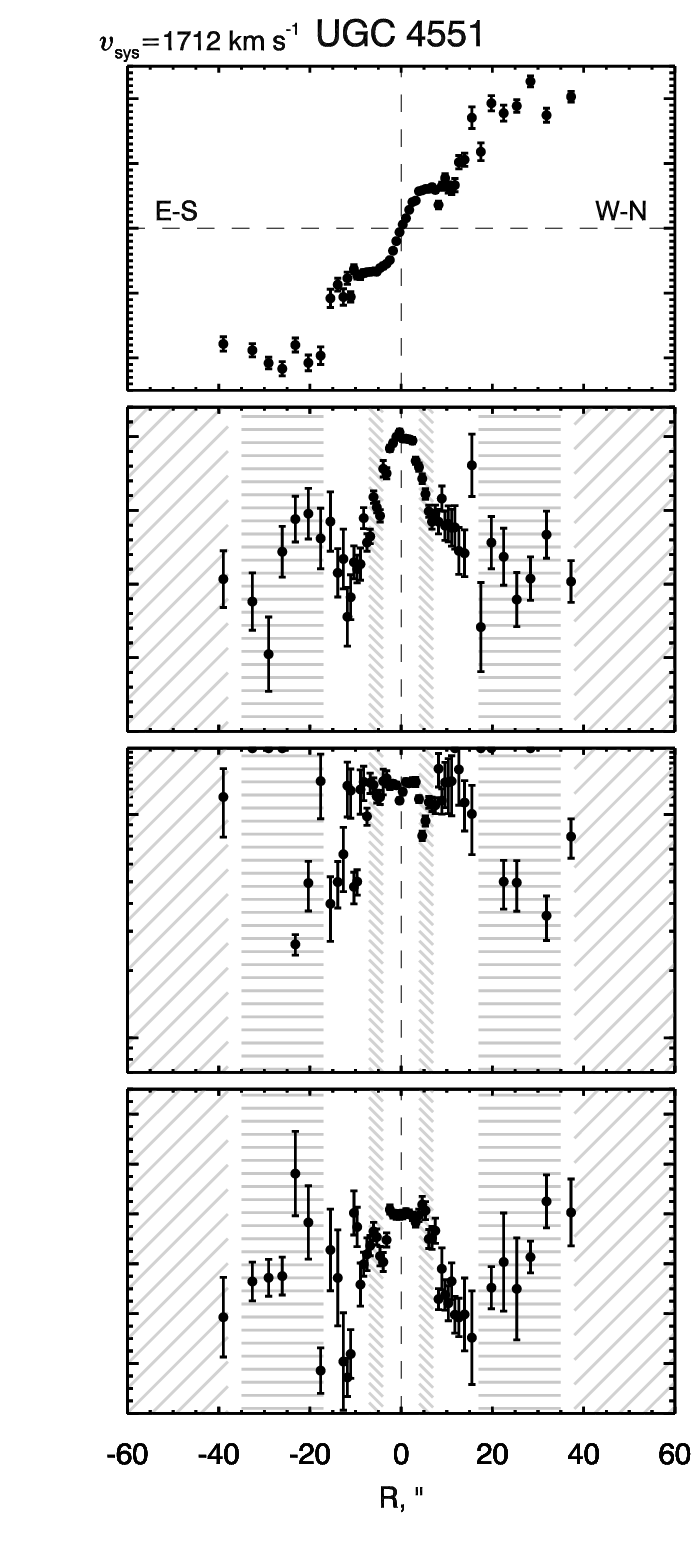}
  \hspace{-10mm}
  \includegraphics[width=0.36\textwidth]{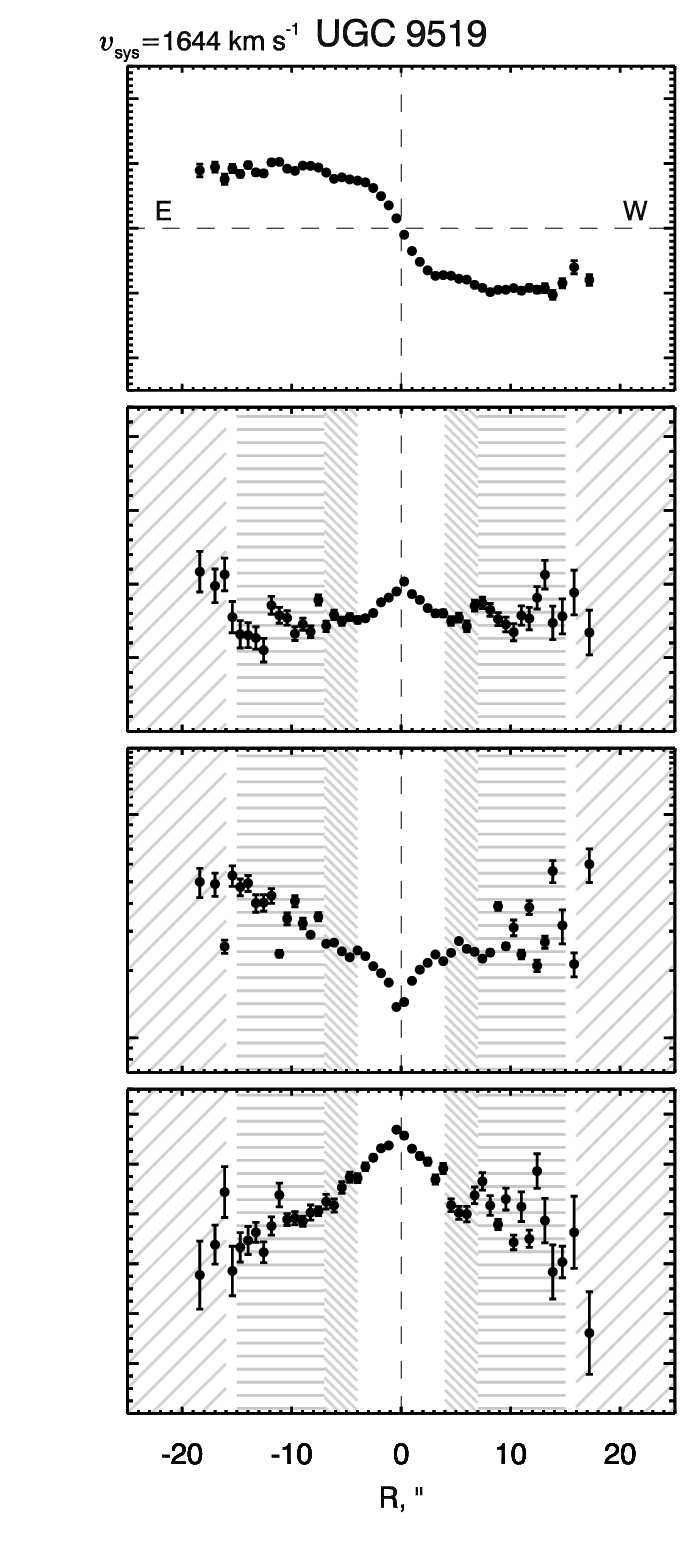}
} \addtocounter{figure}{-1}
 \caption{(Cont.)}
\end{figure*}

\section{RESULTS}\label{results:Katkov_n}

Fig.~\ref{kin_stpop_profiles:Katkov_n} presents the derived radial profiles of the
stellar component properties for each studied galaxy: the line-of-sight velocity profile 
of stars~$v$, stellar velocity dispersion $\sigma$, and the properties of the stellar populations: 
the SSP-equivalent age~$T$ and the SSP-equivalent metallicity~[Z/H]. In addition, we have analyzed 
the available photometric data for these galaxies and have identified the areas where the disc starts
to dominate in the total surface brightness, where the more complex morphological structures such
as rings and lenses are present, and where in the center of a galaxy the region of the bulge prevalence 
can be distinguished (while the nucleus-dominated region is excluded). For this purpose, we have used 
data from the public archive of the SDSS survey, DR9, in the $r$ filter (for the most
galaxies), or, if some galaxies are not observed in the SDSS survey, we have used the 2MASS survey data, 
summed in the \mbox{$J$,~$H$,~$K$} filters (for NGC\,2350, NGC\,6798, IC\,1502), and also the white-light
images obtained with the \mbox{SCORPIO-2} in the imaging mode during
the spectral observations of NGC\,6798 and IC\,1502. For the NGC\,6798
and IC\,1502 the image-analysis results have coincided in the near
infrared bands and in the white light. For every galaxy we have fulfilled an isophote analysis 
and then have treated the profiles of azimuthally averaged surface brightness. The outer regions
being well described by an exponential law of the surface brightness decrease outward where also
constant ellipticity of the isophotes is observed have been considered as the disc dominance regions. 
The local brightness excesses are sometimes visible in the surface-brightness profiles, which we 
identify as rings or lenses. To estimate the bulge parameters, we have used a fixed range of
radii, \mbox{$4\arcsec$--$7\arcsec$}. Over the radial ranges restricted in this way, we have made 
the averaging of the stellar population parameters taking the weights inversely proportional to the
squares of the individual estimate errors, \mbox{$w_i=1/\delta p_i^2$}. The mean parameter estimates 
for the bulges, discs, and rings or lenses are listed in Table~\ref{table_stpop:Katkov_n}. 
A comparison of the derived parameters is also plotted graphically \mbox{in
Figs.~\ref{plot_bulge_disc:Katkov_n}, \ref{plot_mgfe:Katkov_n},
and \ref{plot_sig:Katkov_n}.}

\subsection{IC\,875}

In the HyperLeda and NED databases this galaxy is classified as a lenticular, and hence we have included it 
into our sample. However, after the analysis of the spectral data, we have found that it proved to possess 
rather slow rotation. If we adopt the inclination as \mbox{$i=50\degr$} from the NED, then 
\mbox{$V_{\rm rot}=\Delta V_{\rm LOS}/\sin i \approx 65$~km\,s$^{-1}$}, while the stellar velocity dispersion is
measured to be \mbox{$\sigma \approx 110$~km\,s$^{-1}$}, that gives the ratio of the rotation velocity to the 
velocity dispersion \mbox{$V_{\rm max}/\sigma \approx 0.6$}. The ratio of the major
and minor axes, taken from NED, gives the isophote ellipticity of  \mbox{$\epsilon = 0.25$}. 
These values put the galaxy at the Binney's diagram (\mbox{$V_{\rm max}/\sigma$--$\epsilon$}) 
onto the line of isotropic spheroids supported by rotation; this dynamical status is also common for
low-luminosity elliptical  galaxies~\citep{davies_1983}. So we conclude that IC\,875 is probably an elliptical galaxy. 
We hence eliminated it from our sample of isolated lenticular galaxies, and it was not used to construct
diagrams and distributions over the parameters for the isolated S0~galaxies.

\subsection{IC\,1502}

In the NED, this galaxy is classified as S0$^+$. It has a rather large inclination  \mbox{$i=64\degr$,} 
according to the HyperLeda. The radial line-of-sight velocity profile reveals regular rotation. The main 
feature of this galaxy is that having the solar metallicity, the SSP-equivalent age of the stars is about
\mbox{15--17~Gyr} over the whole measured radial range. The fact that this age has appeared larger than 
the cosmological age of the Universe should not be confusing. The main input components of the
evolutionary synthesis models of the stellar populations are isochrones of stars, which come from the
stellar evolution theory and from the empirical library of stellar spectra. It should be borne in mind 
that timescales of the stellar evolution used so far have nothing in common with the cosmological models
of the Universe and provide absolutely independent channel of information about the age of the dominant 
stellar populations in galaxies. The radial profiles of the stellar parameters in IC\,1502 demonstrate the central
kinematically detached region, which betrays itself as a steep stellar velocity gradient and a small decrease 
in the stellar velocity dispersion. The metallicity profile reveals a marginal dip. All the combination of the
features is probably the evidence of evolutionary isolation of the nuclear star cluster, although as for the 
age of the nuclear stellar population it does not stand away within the measurement error.

\subsection{NGC\,16}

This galaxy has a small bar aligned at a right angle to the line of nodes
of the galactic disc and a faint late-type companion, with the magnitude difference toward the host 
of $\Delta m_B=4^{\rm m}$,~\citep{barway_2005} at 500 kpc from it. The isolation index between the companion 
and the host galaxy is $II=1.9$, but since the mass difference is about two orders of magnitude, 
we consider NGC\,16 as a quite isolated from the external tidal effects. The galaxy has an extended 
rigid-body rise of the rotation curve of up to $10''$ from the center, where the rotation velocity 
reaches a plateau, and its stellar disc is rather dynamically hot, \mbox{$\sigma >100$~km\,s$^{-1}$}. 
The stellar component in NGC\,16 has an intermediate age, the bulge and the disc being of similar ages
in the mean, although individual age estimates in the bins demonstrate large scatter.

\subsection{NGC\,2350}

This poorly studied galaxy was unfortunately missed by the SDSS survey. According to the HyperLeda and NED, 
it is classified as S0/a. In the white-light image obtained with the SCORPIO-2 at the 6m Russian telescope, 
we see a complex distribution of surface brightness shaped as bright spots at the edges of the bar -- the so-called
``ansae'' phenomenon. In addition, from the analysis of the emission-line component of the spectrum, 
the galaxy has an extended disc of ionized warm gas. Plotting the emission-line ratios onto the diagnostic 
BPT-diagrams, we have made sure that it is excited mainly by the radiation of young stars \citep{ilg_gas}; 
i.e., this lenticular galaxy is undergoing current star formation over the entire disc. Correspondingly, 
the averaged age of the stars within the area bordered  by the ``ansae'' is pretty young:
individual estimates in the bins oscillate between 1 and~2~Gyr. In the central region of the galaxy 
the stars demonstrate a very low metallicity. This may be due to a previous event of a metal-poor 
companion infall. At the same time this region distinguishes itself in the rotation curve and at the
velocity dispersion profile, but, to our surprise, it does not stand out at the age profile. According 
to our estimates, the metallicity of the gas is only slightly subsolar even in the regions with the very
low stellar metallicity. Therefore, the ionized gas in this galaxy is probably not related by its origin 
to a possible merger having produced the nuclear stellar cluster.

\subsection{NGC\,3098}

It is a well-known isolated, edge-on lenticular galaxy. Its photometric structure was studied
by~\citep{seifert_scorza}. It was noted that a bulge of the galaxy was small and compact, and 
between the bulge and the disc, approximately at the radius of~15\arcsec, there was a ring
of enhanced stellar surface brightness. Our results (Fig.~\ref{kin_stpop_profiles:Katkov_n}) 
confirm the small effect of the bulge---the   rotation velocity growth in  the center of the galaxy 
is very slow---and demonstrate a homogeneous age of 5--8~billion years for the stellar populations  
across the whole galaxy as well as a sharp drop of the stellar metallicity in the transition from the
central part of the galaxy to the region dominated by the large-scale  stellar disc. The stellar disc
looks dynamically cold, $\sigma<60$~km\,s$^{-1}$.

\subsection{NGC\,3248}

Beyond a radial distance of 100~kpc, the galaxy is surrounded by a
dozen of faint companions, the brightest of which is only $3^{\rm
m}$ fainter than the host galaxy. The central region of the
galaxy, $R<15$\arcsec, has been investigated in detail by means of panoramic
spectroscopy with the SAURON spectrograph within the ATLAS-3D
survey \citep{cappellari_atlas7_2011,davis_2011}.
The galaxy is found to contain a lot of gas, ionized, neutral and
molecular, and all this gas counterrotates with respect to the
stellar component. We have expanded the kinematic profiles to the
distance of 30\arcsec\ from the center  and confirmed the
counterrotation of the gas concentrated in the central region of the
galaxy. The ionized-gas excitation is shock-like, and there are no signs of
ongoing star formation. The age of the stellar populations both in
the center and in the galactic disc is intermediate, but the
central regions are significantly more metal-rich than the disc.

\begin{figure*}[!ht]
\centerline{
  \includegraphics[width=0.34\textwidth, bb= 50 358 320 593,clip]{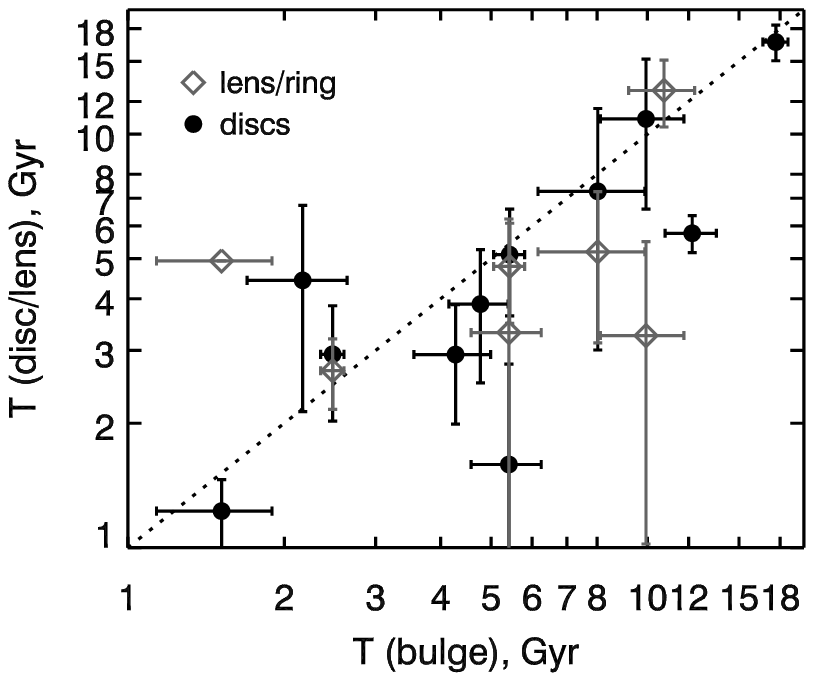}
  \includegraphics[width=0.33\textwidth]{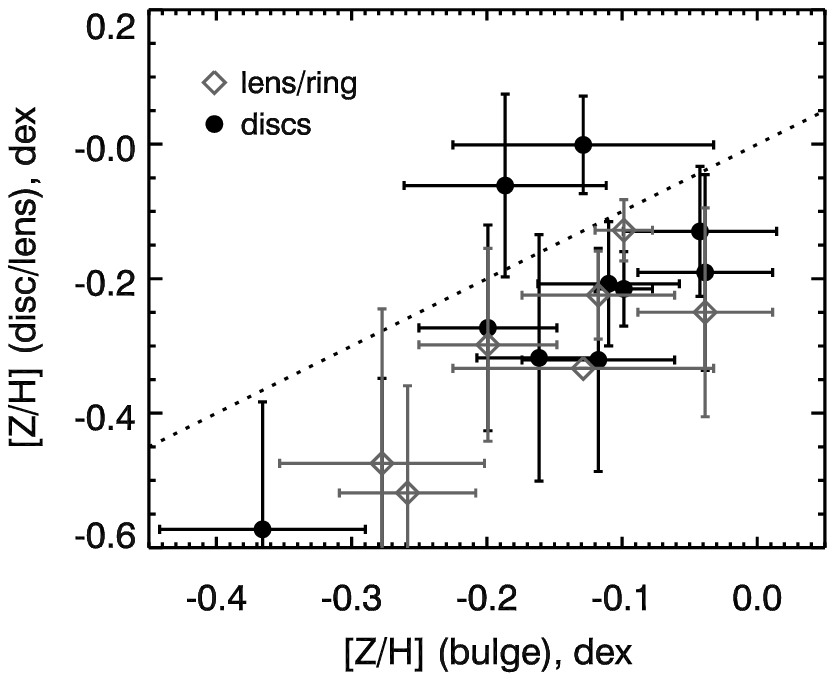}
  \includegraphics[width=0.33\textwidth]{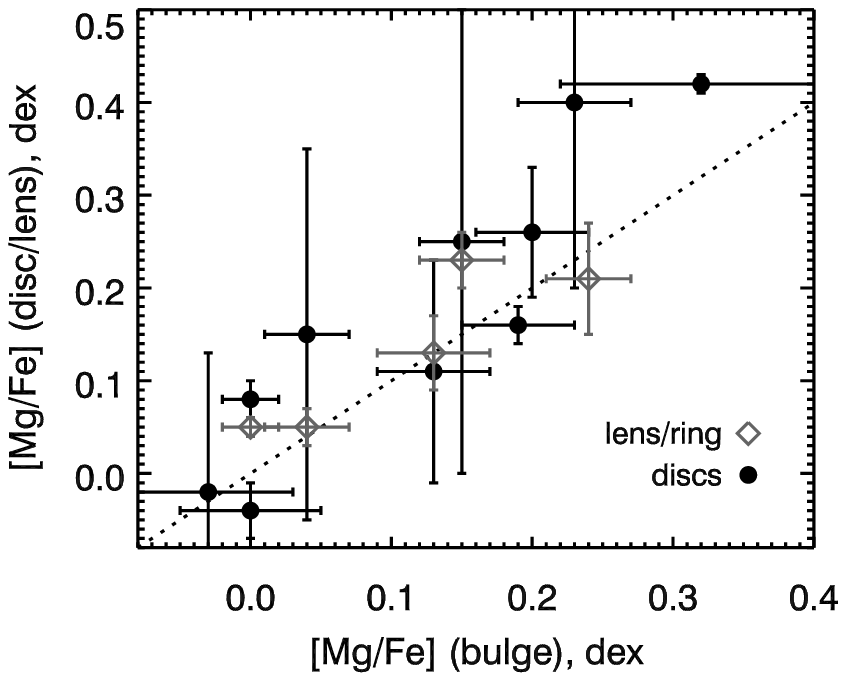}}
\caption{A comparison of parameters of the stellar populations in
the bulges with the properties in the discs and the disc components:
lenses/rings. The dashed line marks the line of equality.}
\label{plot_bulge_disc:Katkov_n}
\end{figure*}

\begin{figure*}[!ht]
\centerline{
  \includegraphics[width=0.34\textwidth, bb= 50 358 320 593,clip]{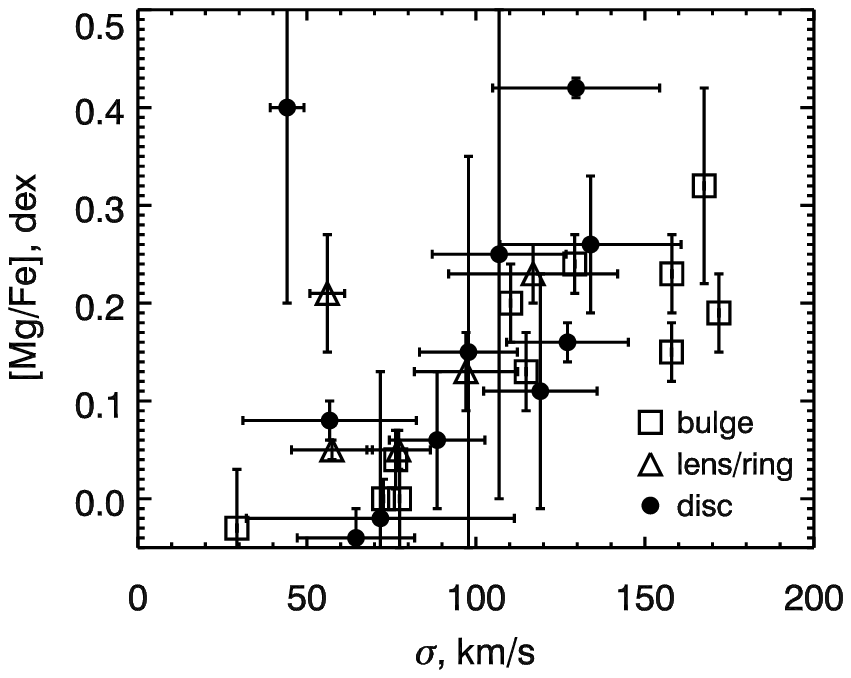}
  \includegraphics[width=0.34\textwidth, bb= 50 358 320 593,clip]{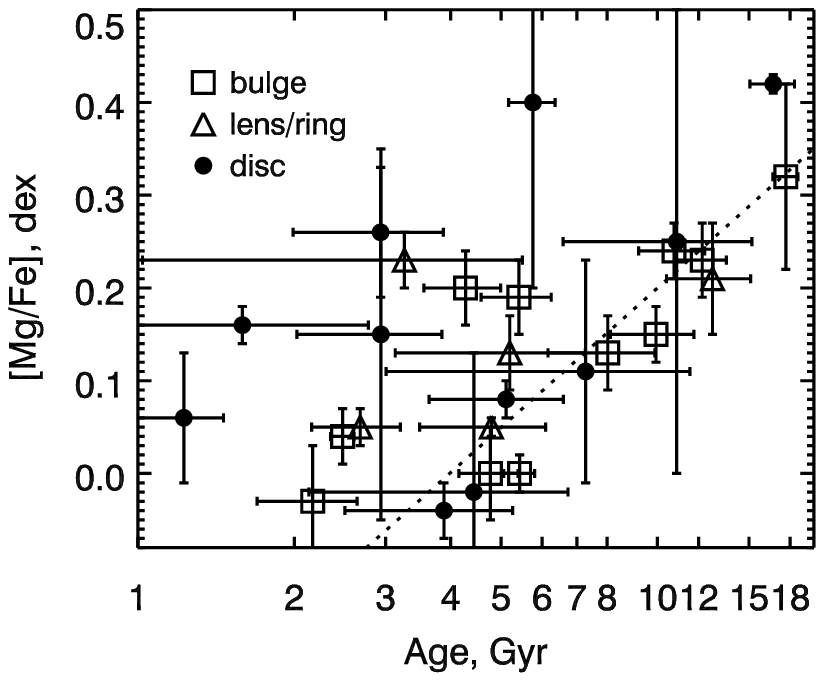}
  \includegraphics[width=0.33\textwidth]{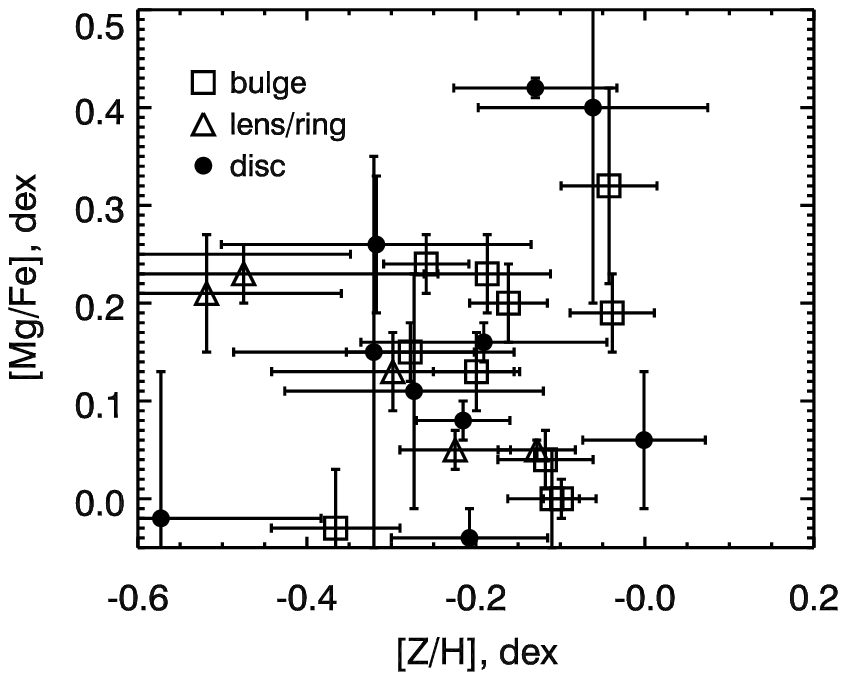}
}
 \caption{A comparison of the velocity dispersion, ages, and metallicities with the  relative  
enrichment of the $\alpha$ elements ([Mg/Fe]) for different structural components of the galaxies.}
\label{plot_mgfe:Katkov_n}
\end{figure*}

\subsection{NGC\,6615}

This galaxy is classified as barred in all the extragalactic databases. 
Visual inspection of the SDSS images confirms that a compact bar is present 
and oriented almost orthogonally to the major axis of the galactic disc. The surface 
brightness profile reveals a noticeable extended stellar lens with a flat brightness 
distribution, while the exponential disc itself starts at the radii greater than
40\arcsec. The sensitivity of our spectroscopy with the high-resolution VPHG2300 holographic 
grating (the grism) was not enough to reach the disc, and we have identified the characteristics
of the stellar populations only in the bulge and in the lens. The age of the stellar population 
is uniformly old throughout the entire studied part of the galaxy. It is possible that between
the bulge/bar and the lens a relatively young narrow stellar ring is present. The metallicity 
is below the solar everywhere, while in the lens it is significantly lower than solar, at least 
by a factor of 3--4. The emission-line gas in the galaxy is not detected, but the stellar lens is
dynamically cold.

\subsection{NGC\,6654}

It is a luminous galaxy with a large-scale bar and a large-scale stellar disc 
of low surface brightness. The NED database gives it a classification as
(R$'$)SB(s)0/a. We have accepted the main parameters of the photometric
structure of the galaxy from \citep{laurikainen_2010}. The outer edge of the disc 
reveals an emission-line gaseous ring with ongoing star formation. Apart from this 
outer ring, the ionized gas is only present in the central region of the galaxy, 
where it demonstrates shock-like excitation and rotates too fast to be in the main symmetry  
plane of the galaxy which is inclined by some $45\degr$ to the line of sight
(according to the HyperLeda). In our work \citep{ilg_gas} we speculated that the  
central gas rotates in the plane inclined to the main symmetry plane of the galaxy. 
The characteristics of the stellar population
(Fig.~\ref{kin_stpop_profiles:Katkov_n}) indicate that the disc is
younger and richer by metals than the bulge. Keeping in mind the
residual star formation at the periphery of the disc, we can conclude
that NGC\,6654 is a rare case of a lenticular galaxy, where rejuvenation 
(or secondary star formation) has taken place in the disc and not in the central region. 
Moreover, given the noticeable overabundance of the magnesium relative to the iron in the 
stellar disc (Table~3), this rejuvenation has had character of a very brief starburst
provoked by expanding density wave.

\begin{figure*}[!ht]
\centerline{
  \includegraphics[width=0.4\textwidth, bb= 50 358 320 593,clip]{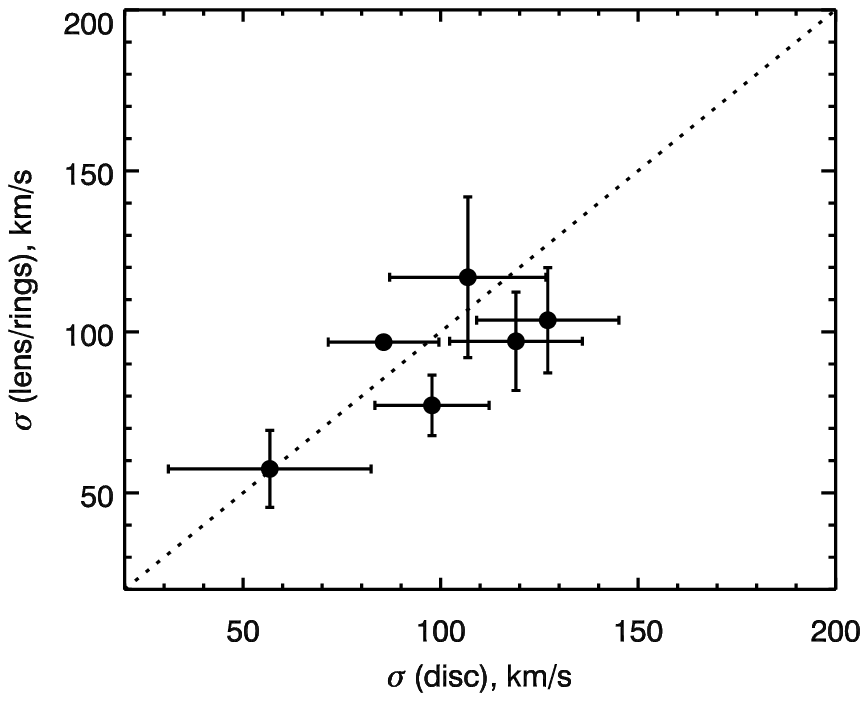}
\hspace{5mm}
  \includegraphics[width=0.4\textwidth, bb= 50 358 320 593,clip]{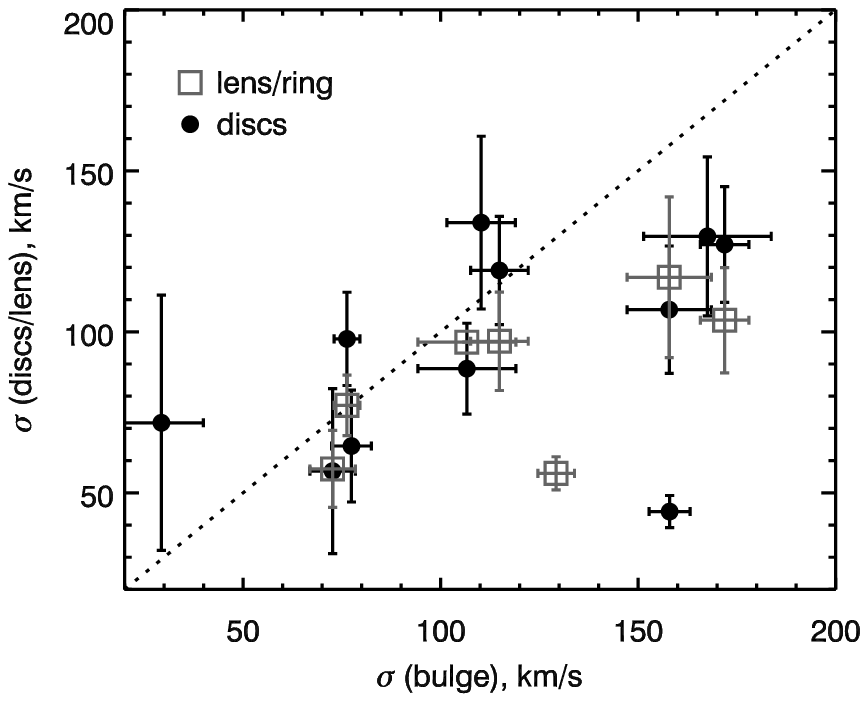}
}
 \caption{Left: a comparison of the stellar velocity
dispersion   in the disc and in separate structures---rings or
lenses. Right: a comparison of the stellar velocity dispersion in
the bulge and in the disc structures.} \label{plot_sig:Katkov_n}
\end{figure*}

\subsection{NGC\,6798}

It is another galaxy which was studied earlier within the 
\mbox{ATLAS-3D} survey~\citep{cappellari_atlas7_2011,davis_2011}, like NGC\,3248,
and where a counterrotating disc of the ionized gas was found, while the neutral hydrogen  
in this galaxy is extended far beyond the stellar disc. Analyzing the photometric structure 
of the galaxy with the white-light image obtained by using the \mbox{SCORPIO-2}, we have noted the
presence of a stellar ring at the radius of \mbox{15\arcsec--20\arcsec}. This ring is \mbox{$5\pm2$~Gyr} 
old that, taking into account the error bar, almost coincides with the age of the disc of
\mbox{$7\pm4$~Gyr }. However the ring is slightly cooler dynamically than the disc that is consistent with its
perhaps younger age. At the same time the metallicity of the stars throughout the galaxy is homogeneous, 
being half-solar.

\subsection{NGC\,7351}

It is a dwarf lenticular galaxy with a low stellar velocity dispersion 
both in the center and in the disc, and with a slow rotation
velocity. Its stellar kinematics was previously investigated by~\citep{simien_prugniel}. 
As it might be expected in the low-luminosity galaxy, the metallicity of its stars is below
the solar one, but while in the center it is lower than the solar only by a factor of
2.5, in the disc it is lower at least by a factor of 4. The ages of the stellar
population monotonically increase  from the center to the edge of the galactic disc, 
from 1.5~Gyr in the nucleus to \mbox{5--8}~Gyr in the disc. The comparison with the characteristics 
of the gaseous component which we published in \citep{ilg_gas} has shown that in
the center the ionized gas rotates together with the stars and is excited by 
current star formation, while outside the central region the gas distribution
deviates from the plane of the galactic disc, and its kinematics becomes
strongly decoupled from the stellar kinematics. In this case the accretion of external gas 
has obviously taken place from a highly inclined orbit, and due to this inclination stationary 
conditions for star formation (rejuvenation) have emerged only in the center of the galaxy,
where the gas has accumulated and come to the galactic symmetry plane.

\subsection{UGC\,4551}

It is one more galaxy with a counterrotating gaseous component, which may
rotate towards the stars right in the plane of the galactic disc \citep{ilg_gas}. 
Nevertheless, both the central part of the galaxy and its external disc contain uniformly
old, \mbox{$T>10$~Gyr}, stellar populations (Fig.~\ref{kin_stpop_profiles:Katkov_n}).  
A stellar lens with a flat brightness profile is observed between the bulge and the disc.
Here, in the lens, stellar population is significantly younger than in the disc and in the bulge. 
From the dynamical point of view both the lens and the disc look rather ``hot'' 
\mbox{$\sigma >100$~km\,s$^{-1}$}.

\subsection{UGC\,9519}

This galaxy, although having an almost face-on orientation with respect to the
line of sight ($i=23\degr$, HyperLeda), demonstrates very rapid apparent rotation of the stars
(Fig.~\ref{kin_stpop_profiles:Katkov_n}).
We \citep{ilg_gas} noted earlier strong divergency of the kinematics of the stars and 
ionized gas in our long-slit data. Panoramic spectral data of the ATLAS-3D
survey \citep{cappellari_atlas7_2011,davis_2011} gives rather evidence for the inner  
polar gaseous disc in this galaxy. As in the case of NGC\,7351, the SSP-equivalent age of 
the stars monotonically rises along the radius of the galaxy, from
1.5~Gyr in the nucleus to about~5~Gyr in the disc. The disc, being the oldest stellar component 
of the galaxy, is also the most dynamically hot, $\sigma \approx 100$~km\,s$^{-1}$. 
It is also metal-poor, about twice poorer than the solar chemical composition. In the center 
of the galaxy the metallicity gets slightly higher than the solar one, that may be understood 
given recent additional star formation activity here.

\section{DISCUSSION}\label{discussion:Katkov_n}

Here we wish to compare the stellar population properties of the
different structural components of the galaxies: the comparison of the mean ages
will help to trace a sequence of basic evolutionary stages in the life of a galaxy, 
while the magnesium-to-iron abundance ratio will allow to estimate the typical duration
of the main events of star formation in various components of a galaxy. The main cosmological
paradigm to date, the LCDM  model of the evolution of the Universe, predicts that the classical 
spheroidal bulges of the early-type disc galaxies are the first to form in a major
merging event, and only after their shaping the large-scale discs have to emerge around them 
by a smooth accretion of cold gas and subsequent star formation. However, there are many data 
which contradict these predictions: again and again, the imaging surveys of galaxies
demonstrate a correlation between the metric scales of the bulges and the discs. Interestingly, 
this correlation is always found independently on whether the given sample of
galaxies is populated by classical or pseudobulges~\citep{macarthur03,mendez_abreu,laurikainen_2010}.
Thus, it looks like the formations of bulges and discs in galaxies are synchronized.

This is exactly what we see in Fig.~\ref{plot_bulge_disc:Katkov_n}. The comparison of the mean
ages of the stellar populations in the bulges and in the flat components, namely, in the discs and lenses, 
shows that our galaxies are concentrated to the bisector, i.e., on the average the ages of the stellar
populations in the bulges and in the discs are the same. There are only two objects in which the stellar age 
of the disc is certainly less than the stellar age of the bulge. This is a particular feature as 
concerning the evolution of our {\it isolated} lenticular galaxies: at the same diagram for the S0 galaxies 
in denser environments, studied by \citet{sil_s0}, the objects are assembled
in the left upper corner, above the bisector, i.e., the ages of the discs are typically either equal 
or older compared with the bulge stellar ages. We expected such difference between the S0
galaxies in groups and in isolation: all the known mechanisms, both gravitational and gas-dynamical, 
of the external influence on a galaxy are associated with dense environment and lead to 
``inflow'' of the gas into galactic center, thus provoking a secondary star formation burst 
exactly in the central areas, in the regions dominated by the bulges \citep{bekki_couch,kronberger}.
Interestingly, the bulges and discs have appeared to possess equal magnesium-to-iron abundance ratios. 
This means that either the star formation ends quickly in both structures or that it goes on for billions 
of years here and there. This result, namely, the similarity of both the mean ages and the Mg/Fe ratios 
in the bulges and discs allows us to strengthen the conclusion on the synchronous formation of bulges 
and discs: the star formation in both components starts and ends quasi-simultaneously. Despite this, 
the mean metallicity of stars in the discs is significantly lower than that in the bulges. Does that 
mean that the accretion of metal-poor gas occurs  mainly to the external discs, while the ``fuel'' for 
star formation in the bulges comes there after its enrichment by the heavy elements in the discs?

Fig.~\ref{plot_mgfe:Katkov_n} compares Mg/Fe ratio,
characterizing the duration of the main starforming episode,
with the other characteristics of the stellar populations for all
three types of structural components. Again, we see an impressive
synchronizing of the evolution of the bulges and the discs: through all the
relations, the signs marking the different structural components are uniformly 
mixed in the graphs. The correlation of the stellar magnesium-to-iron ratio with the 
stellar velocity dispersion, characterizing the local gravitational potential, 
is well-known among elliptical galaxies and bulges \citep{trager} and is considered 
to be a proof of connection between the star formation efficiency and the
potential well depth. However, the left panel of Fig.~\ref{plot_mgfe:Katkov_n} 
reveals for the first time that this correlation takes also place among the discs 
of isolated lenticular galaxies, and it looks even tighter than the one for the
bulges. By analogy, we can suppose that the deeper is  the
local potential well in the galactic plane, the higher is the local
accretion rate of the external gas, and the higher accretion
rate provides more efficient star formation. The middle
plot of Fig.~\ref{plot_mgfe:Katkov_n} compares the Mg/Fe ratios
and the ages of the stellar systems. We can see the linear envelope on
the right (the dotted line), where most of our galaxies concentrate.
These are probably the stellar systems that have begun their star
formation together in the early Universe, at $z=2$--$3$, and have finished it differently. 
Those that had finished it quickly have a higher average age of the stellar populations 
and a higher Mg/Fe ratio, while those where the star formation continued for many
billions of years evolved to the solar Mg/Fe ratio. However, the point distribution in the middle 
panel of Fig.~\ref{plot_mgfe:Katkov_n} has an amorphous tail to the left of
the dotted line, and this tail comprises both the discs and the bulges. These are stellar 
systems where the last star formation episode took place {\it later} than commonly in the
studied population of galaxies. Indeed, in order to have both [Mg/Fe]\,$=+0.2$ and the average 
stellar population age of 1.5--3~Gyr, they needed to undergo their only 1.0--1.5~Gyr of active
star formation much later than at the redshift of $z=0.5$. Therefore, it turns out that the star 
formation episodes in the structural components -- the discs and the bulges -- of isolated lenticular 
galaxies might take place at different times and have various durations.

And finally, a few words about the lenses in the S0-galaxies, which are usually thought 
to be genetically related with former starforming rings and dissolved bars. Until now, 
the common point of view was that stellar populations in lenses are old and that the lenses 
are dynamically hot \citep{kormendy_n1553,lauri13}. However, this point of view is based on a 
study of a few objects. In our small sample we were able to determine the properties of the stellar
populations in seven lenses and old rings. We can certainly conclude that the lenses we have studied 
represent disc structures. On average, they have the same velocity dispersions as the surrounding discs 
(the left panel of Fig.~\ref{plot_sig:Katkov_n}). When we look at the individual velocity dispersion 
profiles, some local peaks can be noted within the area dominated by the lenses, for instance, in
NGC\,3098, NGC\,6798, and UGC\,4551; however, after averaging across the entire lens-dominated region,
the peaks do not substantially contribute to the resulting averaged lens velocity dispersion, 
because the peaks surpass the velocity dispersion in the disc by no more than \mbox{20--30~km\,s$^{-1}$} 
which is comparable to the typical velocity dispersion measurement error. The characteristics of 
the stellar populations in the lenses/rings are generally the same as in the discs. We have found  
a single lens, located in UGC\,4551, which is considerably younger than the surrounding disc. 
At the same time the lenses of intermediate age, \mbox{3--5~Gyr} old, that do not differ by
their stellar population parameters from the discs, were found in five galaxies. 

Concerning the dynamical status of the central bulges, our small sample of isolated lenticular galaxies 
proved to have practically equal numbers of pseudobulges possessing the same stellar velocity dispersion 
as their discs, and of classical bulges which are dynamically hotter than the discs (the right panel in 
Fig.~\ref{plot_sig:Katkov_n}). This once again confirms that bulges in lenticular galaxies can be
very different if regarding their luminosity and the contribution to the total mass of the galaxy or 
their origin and evolution. And this point is true even if we consider a sample of {\it
isolated} lenticular galaxies, for which the influence of the environment on the evolution seems to be minimized.

As a summary, the results obtained for the stellar components of the sample of isolated 
lenticular galaxies presented in this work have confirmed our long-standing suspicions about 
the influence of the environment's density on the evolution of galaxies. Namely, for isolated 
lenticular galaxies, unlike what is observed for the members of groups and clusters,
there is no fixed epoch when the structural components form: they may start their star formation 
in the bulges or in the discs at a $z> 2$ as well as only a billion years ago. The scatter of 
the mean (SSP-equivalent) stellar ages in the discs of S0 galaxies indeed increases with decreasing 
density of the environment~\cite{sil_s0} and reaches its maximum among the isolated galaxies.

What may affect the morphological ``fate'' of a disc galaxy residing in rarefied surroundings? 
Why can it prove to become lenticular or spiral to the present epoch?
Everything points to the regime of accretion of the cold external gas, which is known to
fuel star formation in the discs of present spiral galaxies over past billions of years, 
and this regime is likely to be stochastic. A recent study on searching for faint companions of isolated
galaxies \citep{kara11} has revealed an interesting statistical feature: in isolated lenticular galaxies 
the companions have systematically larger relative velocities with respect to their host galaxy than 
the companions of isolated spiral galaxies; moreover, the isolated lenticular galaxies have no companions
whatsoever with the systemic velocity difference of less than 50~km\,s$^{-1}$. Does this kinematical feature 
mean that the companions of lenticular galaxies {\it cannot} be accreted by their hosts in the nearest future, 
whereas for the companions of spiral galaxies the dynamical friction is enough to provide their continuous
accretion? Let us reverse the argument: it is possible that the orbital structure of the group of faint
companions may be any one, and those disc galaxies whose system of the companions is dynamically cold 
can obtain fuel for the star formation in their discs and become spirals, but those which happened
to have a dynamically hot orbital system of companions (or in the present epoch they have already ``dropped'' 
onto themselves all the companions that could be dropped) stay lenticulars. A similar hint was noticed by us 
during the analysis of ionized-gas rotation in the isolated disc galaxies.  In half the isolated 
lenticular galaxies containing ionized gas, its visible rotation is counterwise with respect to the direction 
of the rotation of stars \citep{ilg_gas}. This may mean that the gas was accreted from the orbits isopropically 
inclined to the plane of the main stellar disc; while in spiral galaxies the gas with the ``decoupled''
kinematics is much more rare. This might imply that a steady accretion confined to the main stellar disc plane 
provides stable smooth accumulation of the cold gas suitable for continuous star formation, while the 
gas infall from inclined orbits leads certainly to developing gas turbulence, heating, and so preventing the star
formation ignition. For instance, generation of shock waves during the passage of the potential well of the galactic
disc by a polar-ring gas was described by \citet{wakamatsu93}. Or shock waves may be generated during gas accretion through 
the collision of the external gas with the primary gas of the galaxy, already existing in the disc. This effect, 
depending on the geometry of accretion of the external gas, may also affect the shaping of the morphological type 
of a galaxy.

\section{CONCLUSIONS}\label{conclusions:Katkov_n}

In this paper we have presented the results of long-slit spectroscopy
for a sample of isolated lenticular galaxies. As a result of observations with  
the SCORPIO-2 and SCORPIO reducers of the 6-meter BTA telescope of the SAO RAS, 
we have measured radial profiles of the stellar rotation velocity, stellar velocity dispersion, 
the SSP-equivalent ages and the SSP-equivalent metallicities of stars in 12 objects. One of them, 
IC\,875, has proved to be a low-luminosity elliptical galaxy. We have analyzed the
statistics of the characteristics obtained for the radially resolved stellar populations, by basing on 
the data of the remaining 11 galaxies. The ages of the stellar populations in our sample of isolated
lenticular galaxies cover the full range of values, from 1.5 to 15~Gyr, and, unlike the S0 galaxies in 
denser environments, the isolated galaxies tend to have the same ages of
stars in the bulges and discs; obviously, they had no chance of separate bulge rejuvenation. 
The lenses and rings of enhanced stellar brightnesses, detected on radial brighntness profiles in 
7 of 11 galaxies, usually have the stellar populations and the stellar velocity dispersions
indistinguishable from the properties of the discs. We suggest that probably the shaping of the 
morphological type of disc galaxies in complete isolation critically depends on the possible regimes 
(continuity and geometry) of accretion of external cold gas.

\section{Acknowledgments}
The authors thank Dmitry Makarov for his help
with the database of galaxies of the Local Volume and construction
of the sample of isolated lenticular galaxies, and Nikolay Borisov
for his support with the SCORPIO-2 observations at the  BTA. During
the analysis of the sample galaxies we have used the Lyon--Meudon
extragalactic database (HyperLeda), supported by the LEDA team at
the Lyon Observatory CRAL (France), and the database of
extragalactic  data NASA/IPAC~(NED), managed by the the Jet
Propulsion Laboratory of the California Institute of Technology
under a contract with National Aeronautics and Space
Administration (NASA, USA). As a source of photometric data,  we
have used the public archives of the SDSS-III project ({\tt
http://www.sdss3.org}),
 supported by the Alfred P.~Sloan foundation, the
institutions participating in the  SDSS collaboration, the
National Science Foundation (NSF), and the U.S. Department of
Energy, as well as the data of the Two Micron All Sky Survey
(2MASS),  which was jointly performed by the University of
Massachusetts and the  infrared data analysis center at Caltech
with the financial support from the NASA and NSF. Our study of
isolated lenticular galaxies was supported by the RFBR grants
No.~\mbox{13-02-00059a} and \mbox{12-02-00685a}. I.~Katkov
expresses gratitude to the nonprofit  \mbox{\it Dynasty}
foundation for supporting his research. The observations on the
6-meter  BTA telescope were held with the financial support of the
Ministry of Education and Science of the Russian Federation (state
contracts No.~14.518.11.7070, 16.518.11.7073).

\renewcommand{\baselinestretch}{0.7}
{\small
\bibliographystyle{apj}
\bibliography{ref_katkov}
}

\end{document}